\documentclass[aps,reprint,prb,amsmath,amssymb,groupedaddress]{revtex4-1}
\usepackage[dvipdfmx]{graphicx}
\usepackage{dcolumn}
\usepackage{bm}
\usepackage{mathrsfs}
\usepackage{braket}
\usepackage{empheq}
\usepackage{here}
\usepackage{color}

\makeatletter
\let\MYcaption\@makecaption
\makeatother

\usepackage[subrefformat=parens,font=large,margin=0pt]{subcaption}
\makeatletter
\let\@makecaption\MYcaption
\makeatother

\newcommand{\Par}[2]{{\frac{\partial #1}{\partial #2}}}

\newcommand{\F}[2]{\frac{#1}{#2}}
\begin{document}

\title{Method to Observe Anomaly of Magnetic Susceptibility for Quantum Spin
Systems}
\author{Nobutaka Aiba}
\email[n.aiba@stat.phys.kyushu-u.ac.jp]{}
\author{Kiyohide Nomura}
\email[knomura@stat.phys.kyushu-u.ac.jp]{}
\thanks{}
\affiliation{Department of Physics, Kyushu University, Fukuoka, 819-0395, Japan}

\begin{abstract}
In quantum spin systems, a phase transition is studied from the perspective of magnetization curve and a magnetic susceptibility. We propose a new method for studying the anomaly of magnetic susceptibility $\chi$
that indicates a phase transition.
In addition, we introduce the fourth derivative $A$ of the lowest-energy eigenvalue per site
with respect to magnetization, i.e., the second derivative of $\chi^{-1}$.
To verify the validity of this method, we apply it to an $S=1/2$
XXZ antiferromagnetic chain.
The lowest energy of the chain is calculated by numerical diagonalization.
As a result, the anomalies of $\chi$ and $A$ exist at zero magnetization.
The anomaly of $A$ is easier to observe than that of $\chi$, indicating that
the observation of $A$ is a more efficient method
of evaluating an anomaly than that of $\chi$.
The observation of $A$ reveals an anomaly that is different from the Kosterlitz--Thouless (KT) transition.
Our method is useful in analyzing critical phenomena.
\end{abstract}
\maketitle
\section{Introduction}
In condensed matter physics, phase transitions and their corresponding energy gaps are
an important research subject.
Researching these gaps is necessary for studying the behavior of quantum spin systems.
Bethe showed that an $S=1/2$ XXZ chain system had the characteristic of
the absence of a gap.\cite{13}
Later, Haldane argued that the difference between half-spin 
and integer spin systems involved the gap.\cite{23}

Many researchers have observed the energy gap via the magnetization curve
as a function of the magnetic field.
The magnetic field at zero magnetization is equal to the magnitude of the gap.
However, the method of observing the gap is not appropriate for deciding
whether a spin system is gapless or gapped in numerical calculation;
it is difficult to distinguish a gapless system from one with
a very small energy gap.\cite{32}

Hence, Sakai and Nakano\cite{1,28,29,30}
proposed a method for distinguishing a gapless from a gapped system.
They introduced the magnetic susceptibility and used numerical diagonalization.
They demonstrated that the susceptibility clearly shows the variation of
the energy gap with changing magnetization, in comparison to the magnetization curve.
Subsequently, they found the anomaly of the magnetic susceptibility.
The term `anomaly' refers to a divergence in the thermodynamic limit.
This anomaly usually exhibits a phase transition.

In this paper, we propose a novel method of evaluating an anomaly by investigating
the magnetic susceptibility $\chi$ and the fourth derivative $A$ of the energy
with respect to magnetization.
Few investigations of high-order differentials such as $A$ have been carried out.
We show that our method is appropriate for analysis of
the phase transition, compared with the method using the magnetic susceptibility
$\chi$ alone.
The introduction of $A$ resolves the issue of whether the high-order differential of energy diverges.
As a test case, we apply this method to the $S=1/2$
XXZ antiferromagnetic chain, which shows ferromagnetic phase for $\Delta \le-1$, Tomonaga--Luttinger (TL) phase for $-1 < \Delta \le 1$, and antiferromagnetic phase for $\Delta>1$.
Here, $\Delta$ denotes an anisotropic parameter associated with the $z$ component of
the XXZ antiferromagnetic chain.
The lowest energy up to 26 spins of the chain is calculated
by numerical diagonalization on the basis of the Lanczos algorithm.
Subsequently, we analyze the anomalies of $\chi$ and $A$ to observe the phase transition.
The results demonstrate that
an anomaly of $\chi$ at zero magnetization exists under $\Delta>1$, while
an anomaly of $A$ at zero magnetization is shown for $\Delta>1/2$.
Hence, the anomaly of $A$ is easier to observe than that of $\chi$.

The anomaly of $A$ at $1/2<\Delta<1$ is different from that of $A$ at $\Delta=1$, indicating a Kosterlitz--Thouless (KT) transition.
It is well established that the $-1<\Delta \le1$ region corresponds to the TL phase, in which the scaling dimensions vary continuously with the parameter $\Delta$.\cite{5,22}
In the $\Delta>1$ region, a Neel state appears in which the ground state is doubly degenerate with an energy gap.
Under the Hamiltonian of the U(1) symmetry, the change of scaling dimensions from irrelevant to relevant indicates a KT transition that corresponds to the phase transition at $\Delta=1$ in the $S$$=$$1/2$ XXZ chain. 
In contrast, the scaling dimensions influencing high derivatives such as $A$ remain irrelevant for $-1< \Delta \le1$.\cite{45} Thus, the onset of the anomaly of $A$ at $\Delta = 1/2$ is different from the KT point and does not indicate the phase transition. We refer to $-1<\Delta<1/2$ as TL phase (I) and $1/2<\Delta\le1$ as TL phase (I\hspace{-.1em}I), as the TL phase is divided by the anomaly of $A$.
The scaling dimensions influence the corrections for various quantities such as energies, susceptibility, and high derivatives.\cite{45}
Thus, the anomalies of $\chi$ and $A$ indicate the phase transition and the energy gap. 

The starting point of the anomaly of $A$, i.e., $\Delta=1/2$, corresponds to
$N=2$ supersymmetry (SUSY) from correspondence between the XXZ chain
and the free boson model\cite{36} and Ashkin--Teller model.\cite{40}
Moreover, the results of our computations agree with the exact solutions
under $0\le\Delta<1$.
These findings indicate that the method using $A$ is better than that using $\chi$
for analyzing critical phenomena with phase transitions.

This paper is organized as follows.
In Sec.~I\hspace{-.1em}I, the calculation method of $\chi$ and $A$ is introduced.
In Sec.~I\hspace{-.1em}I\hspace{-.1em}I, we present our numerical results
for the $S=1/2$ XXZ chain. 
In Sec.~I\hspace{-.1em}V, we compare our results with available exact solutions 
to investigate the behavior of $A$.
In Sec.~V, we reveal that the anomaly of $A$ is associated with conformal field theory.
The correction term is discussed from the perspective of the boundary conditions and
dimension.
In Sec.~V\hspace{-.1em}I, the anomaly of $\chi^{-1}$ and $A$ is discussed in detail from
the perspective of size dependence.
Section~V\hspace{-.1em}I\hspace{-.1em}I is the conclusion.

\begin{figure*}[ht]
\begin{minipage}[t]{0.33\linewidth}
 \centering
  \includegraphics[keepaspectratio,scale=0.48]{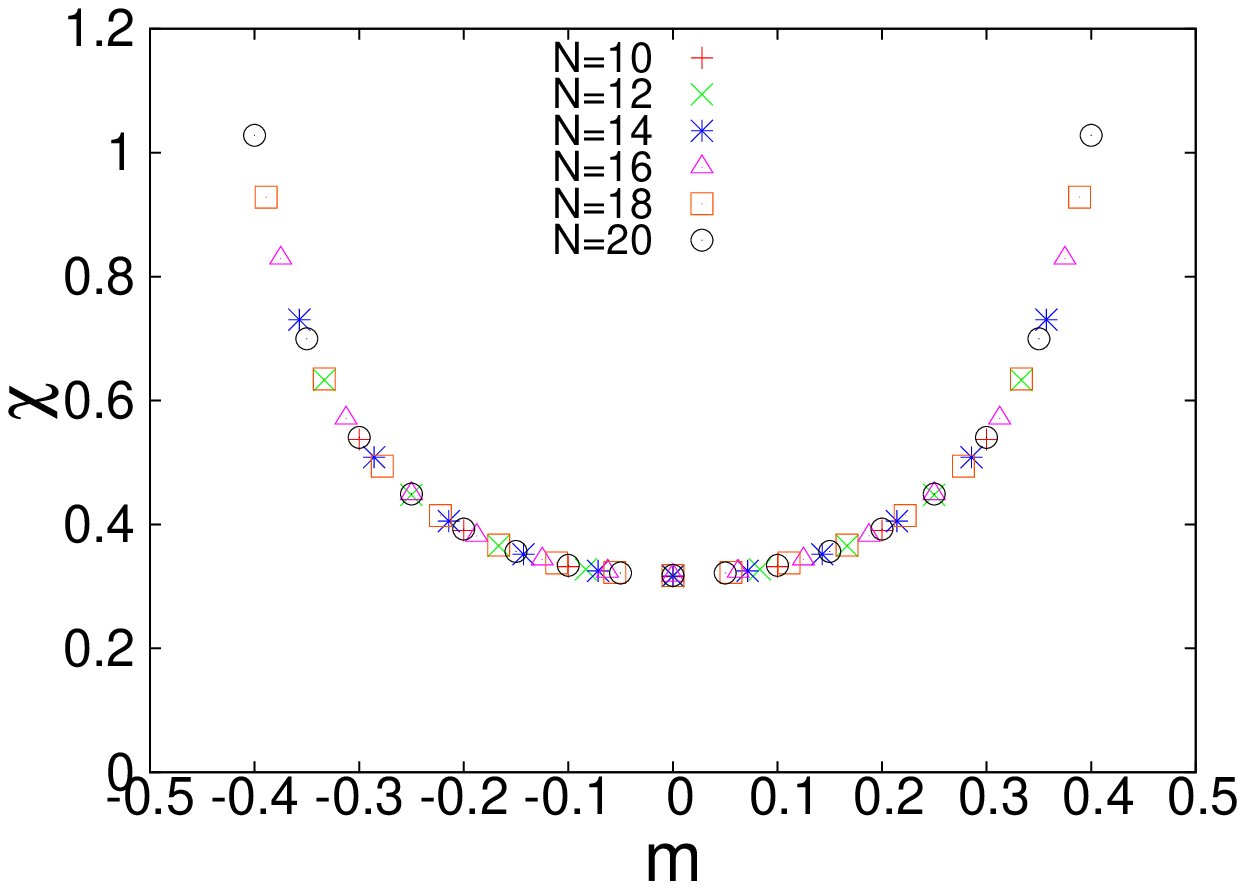}
 \subcaption{$\Delta=0$}
  \label{zika0}
\end{minipage}
\begin{minipage}[t]{0.33\linewidth}
 \centering
  \includegraphics[keepaspectratio,scale=0.48]{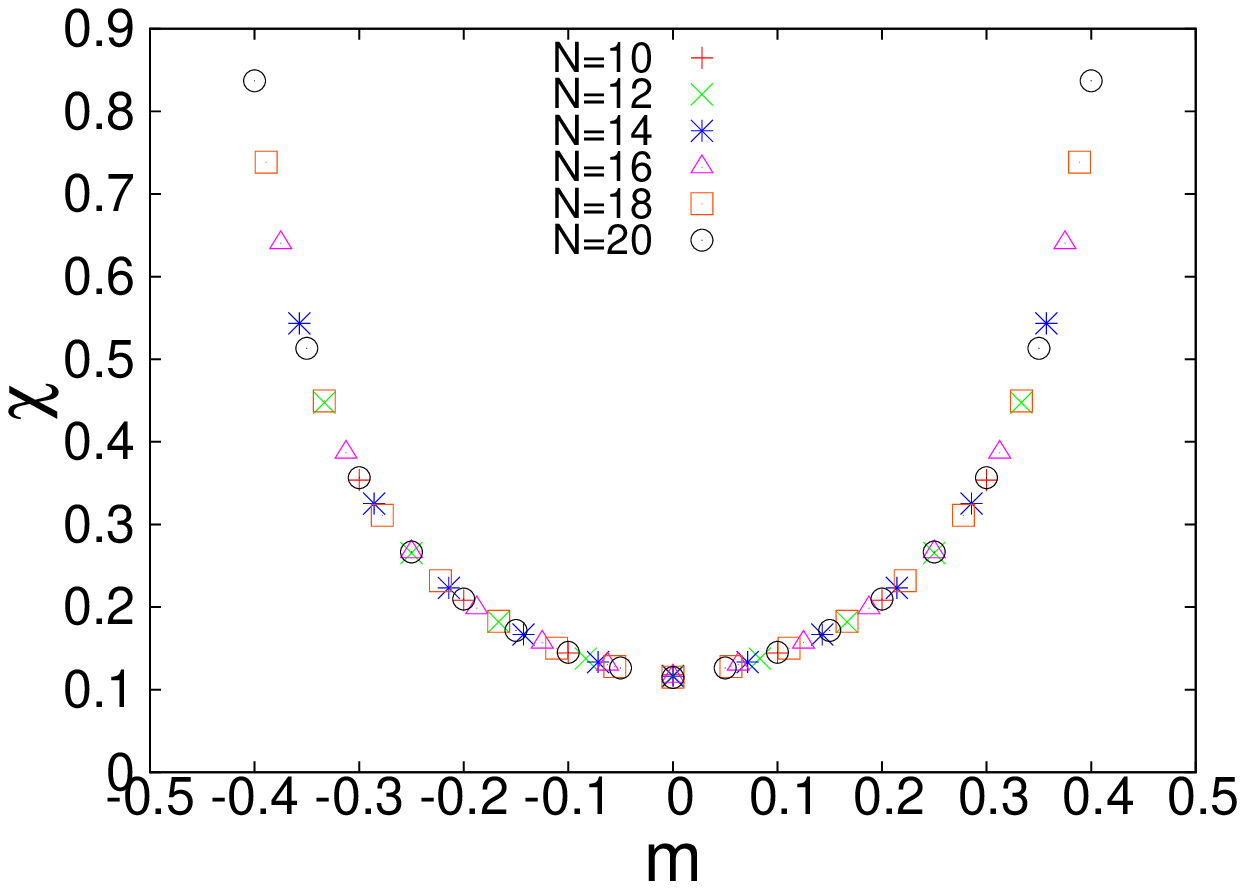}
  \subcaption{$\Delta=1$}
  \label{zika1}
\end{minipage}
\begin{minipage}[t]{0.32\linewidth}
 \centering
  \includegraphics[keepaspectratio,scale=0.48]{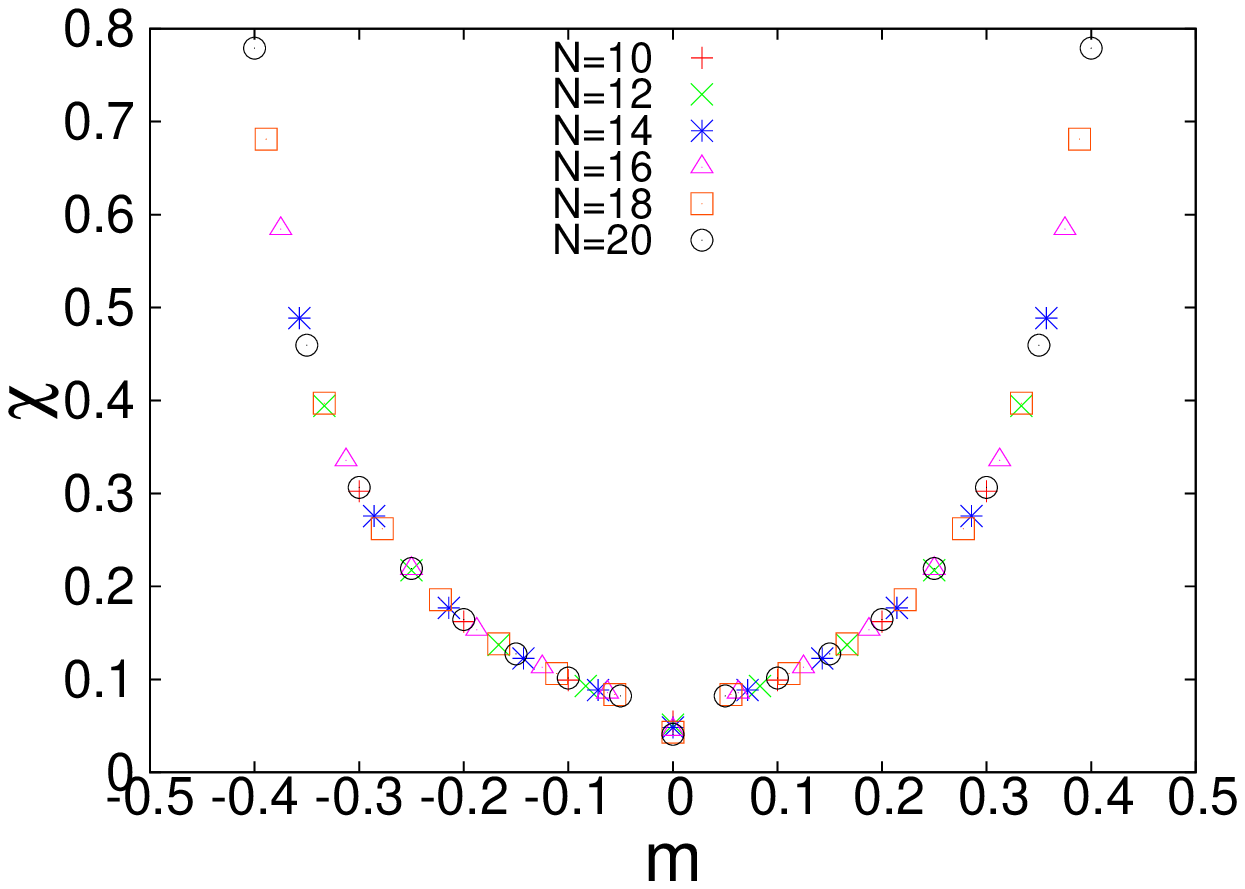}
  \subcaption{$\Delta=2$}
  \label{zika2}
 \end{minipage}
 \caption{
 Magnetization dependence of the magnetic susceptibility $\chi$
 of the $S=1/2$ XXZ antiferromagnetic chain for several system sizes $N$: 10, 12, 14, 16, 18, and 20.
\subref{zika2} shows that $\chi$ has a sharp cusp at zero magnetization.
However, \subref{zika2} does not exhibit an anomaly because the size dependence
is small. Thus, $\chi$ does not have an anomaly.}
 \label{chi}
\end{figure*}

\section{Method: Magnetic Susceptibility $\chi$ and Fourth Derivative $A$}
In this section, we introduce the physical procedure to calculate
the magnetic susceptibility $\chi$ and fourth derivative $A$ of energy as a function of magnetization.
First, we define the total spin operator in the $z$ direction as
\begin{align}
\label{2}
 \hat{S}_{T}^{z} \equiv \sum_{j=1}^{N} \hat{S}_j^z,
\end{align}
where $\hat{S}_{j}^{z}$ is the $j$th site spin operator in the $z$ direction
and $N$ is the system size.
This operator and a Hamiltonian $\mathscr{\hat{H}}$ that shows $U(1)$ symmetry
commute: $[\mathscr{\hat{H}}, \hat{S}_T^z]=0$.
Therefore, the relation is obtained that
\begin{align}
 \mathscr{\hat{H}}\ket{\psi} &= E(N,M)\ket{\psi},\label{4}\\
 \hat{S}_{T}^{z}\ket{\psi} &= M\ket{\psi}\ (M=0,\pm 1,.....,\pm N/2),\label{5}
\end{align} 
where $E(N,M)$ is the lowest-energy eigenvalue,  $M$ is the magnetization, and
$\ket{\psi}$ is the simultaneous eigenstate.
The energy of $\mathscr{\hat{H}}$ per site, $\epsilon(m)$, in the thermodynamic
limit is then written\cite{3}
\begin{align}
 \label{6a}
 \lim_{N \to \infty}\F{E(N,M)}{N} = \epsilon(m),
\end{align}
where $m=M/N$ is the magnetization per site. In finite $N$ cases, it is shown that
\begin{align}
\label{7}
 \F{E(N,M)}{N} = \epsilon(m) + C(N,m),
\end{align}
where $C(N,m)$ is a correction term of a finite size.
Generally, $\epsilon(m)$ is analytic for $m$ in the thermodynamic limit.
The term `analytic' means that the function and high-order differential
are continuous (our study treats the high-order differential up to the fourth derivative). 
$C(N,m)$ satisfies
\begin{align}
\label{8}
 \displaystyle \lim_{N \to \infty} C(N,m) &= 0,\\
 \displaystyle \lim_{N \to \infty} C^{(n)}(N,m) &= 0\ (n\ge1),
\end{align}
where $C^{(n)}(N,m)$ is the $n$th derivative of the correction term
with respect to magnetization.
The correction term depends on the boundary conditions and dimension.

Next, we define the magnetic susceptibility $\chi$ and
fourth derivative $A$ in the form 
\begin{align}
 \chi&\equiv\F{1}{\epsilon''(m)},\label{9}\\
 A&\equiv\Par{^2}{m^2}\chi^{-1}=\Par{^4}{m^4}\epsilon(m)\label{10}.
\end{align}
It is shown that
\begin{align}
 &\epsilon''(N,m) \nonumber\\&\equiv N\{E(N,M+1) - 2E(N,M)
 + E(N,M-1)\}\nonumber\\
 &=\chi^{-1}+C''(N,m)+\F{1}{12N^2}\left(\epsilon^{(4)}(m)+C^{(4)}(N,m) \right)\nonumber\\
 &+\mathcal{O}\left(\F{1}{N^4}\right),\label{nikai}\\
 &\epsilon^{(4)}(N,m)\nonumber\\
 &\equiv N^3\{E(N,M+2)-4E(N,M+1)+6E(N,M)\nonumber\\&-4E(N,M-1)+E(N,M-2) \}
 \nonumber\\
 &= A + C^{(4)}(N,m)+\F{1}{6N^2}\left(\epsilon^{(6)}(m)+C^{(6)}(N,m) \right)\nonumber\\
 &+ \mathcal{O}\left(\F{1}{N^4}\right)\label{yonkai},
\end{align}
where $\epsilon^{(n)}(N,m)$ is the $n$th finite-difference between energies.
$\epsilon(N,m)$ is obtained directly from numerical data in finite systems.
$E(N,M+1) - 2E(N,M)+ E(N,M-1)$ becomes a nonzero constant at large $N$
when finite energy gaps exist at $m$. Similarly, $E(N,M+2)-4E(N,M+1)+6E(N,M)-4E(N,M-1)+E(N,M-2)$ becomes a nonzero constant.
For example, we consider the XXZ model at $m=0$ and $m=\pm 1/N$. For a large anisotropic limit $\Delta \gg 1$ in the Neel region, 
there is an energy gap at $m=0$ and the energies are written in the form 
\begin{align}
E(N,M)= E(N,0)+|M| \Delta E ,
\label{eq:energygap}
\end{align}
where $M$ is the magnetization and $\Delta E$ is an energy gap for a finite system.
For large $N$, substituting Eq. (\ref{eq:energygap}) into Eq.(\ref{nikai}), we obtain
\begin{align}
  \epsilon''(N, m)=
  \begin{cases}
    N(2 \Delta E) & ( m=0) \\
    0 & (m \neq 0)
  \end{cases}.
        \label{eq:chigap}
\end{align}
Similarly, substituting Eq. (\ref{eq:energygap}) into Eq. (\ref{yonkai}), we obtain
\begin{align}
  \epsilon^{(4)}(N, m)=
  \begin{cases}
    N^{3}(- 4 \Delta E) & ( m=0) \\
    N^{3}(2 \Delta E) & ( m=\pm 1/N) \\
    0 & (m \neq 0,\pm 1/N)
  \end{cases}.
      \label{eq:yonkaigap}
\end{align}
However, for a finite anisotropy, there is some interaction between magnons. Thus, the above relations are modified to the forms 
\begin{align}
  \epsilon''(N, m)=
  \begin{cases}
    N(2 \Delta E) & ( m=0) \\
    \mbox{finite} & (m \neq 0)
  \end{cases},
        \label{eq:chigap2}
\end{align}
\begin{align}
  \epsilon^{(4)}(N, m)=
  \begin{cases}
    N^{3}(- 4 \Delta E) & ( m=0) \\
    N^{3}(2 \Delta E) & ( m=\pm 1/N) \\
    \mbox{finite} & (m \neq 0,\pm 1/N)
  \end{cases}.
      \label{eq:yonkaigap2}
\end{align}
This means $\epsilon^{(4)}(N,m)$ is $N^2$ times as large as $\epsilon''(N,m)$.
This fact shows that the anomaly of $A$ appears stronger than that of $\chi^{-1}$ in the thermodynamic limit. Thus, we introduce $A$ for observing an anomaly.

Finally, we consider the case in which $\epsilon(m)$ is not analytic.
$\epsilon(m)$ is not analytic for $m$ 
when $\epsilon''(N,m)$ or $\epsilon^{(4)}(N,m)$
diverges. In the thermodynamic limit, it is given by
\begin{empheq}[left=\empheqlbrace]{alignat=2}
 \lim_{N\rightarrow\infty}\epsilon''(N,m) &= \epsilon''(m)\\
  \quad &\Rightarrow \mbox{$\epsilon(m)$ is analytic,} \nonumber\\
 \lim_{N\rightarrow\infty}\epsilon''(N, m) &= \pm \infty\\
  \quad &\Rightarrow \mbox{$\epsilon(m)$ is not analytic}\nonumber.
\end{empheq}
The same holds for $\epsilon^{(4)}(N,m)$.
The divergence of $\epsilon''(N,m)$ and $\epsilon^{(4)}(N,m)$
is equivalent to the fact that $\chi^{-1}$ and $A$ diverge.

\begin{figure*}[ht]
\begin{minipage}[t]{0.33\linewidth}
 \centering
  \includegraphics[keepaspectratio,scale=0.48]{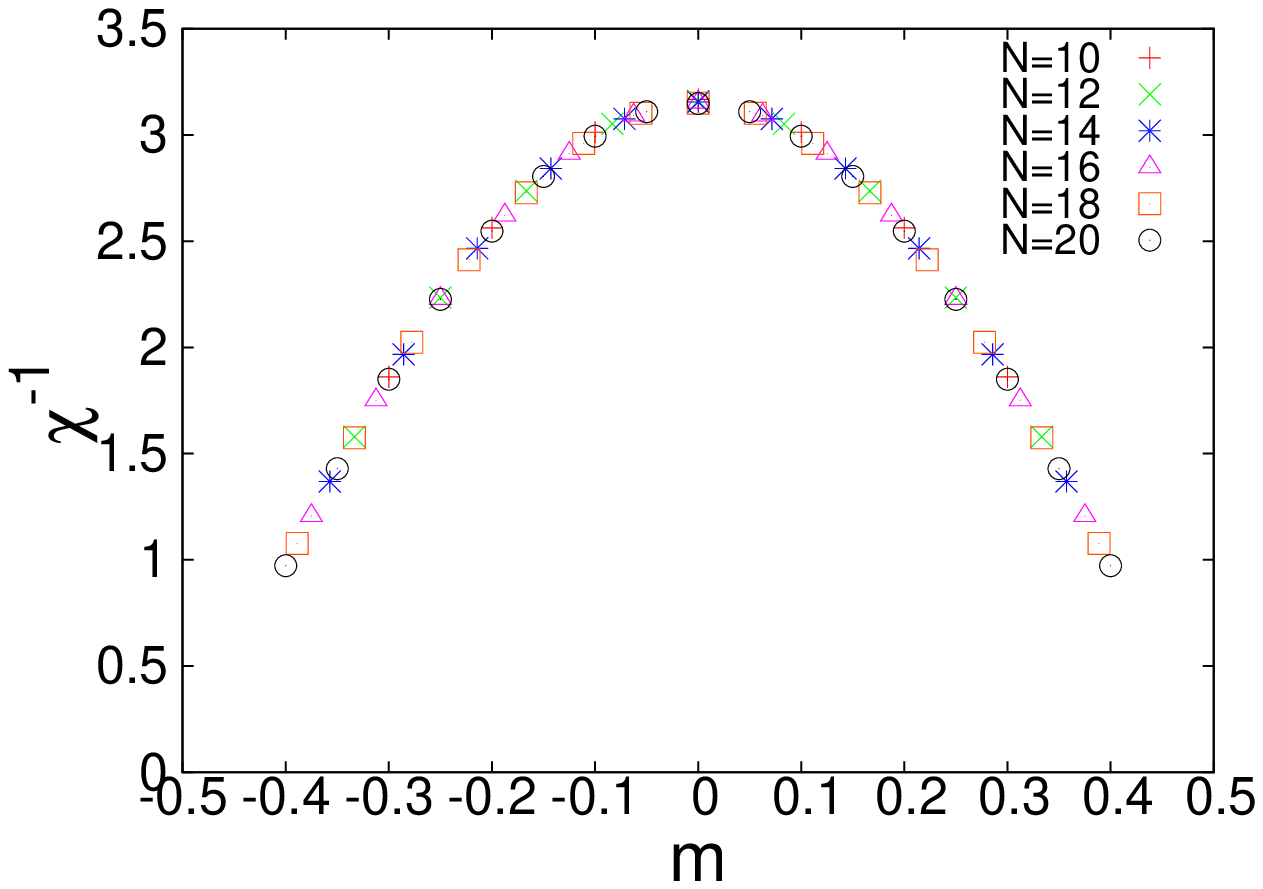}
  \subcaption{$\Delta=0$}
  \label{zikagyaku0}
\end{minipage}
\begin{minipage}[t]{0.33\linewidth}
 \centering
  \includegraphics[keepaspectratio,scale=0.48]{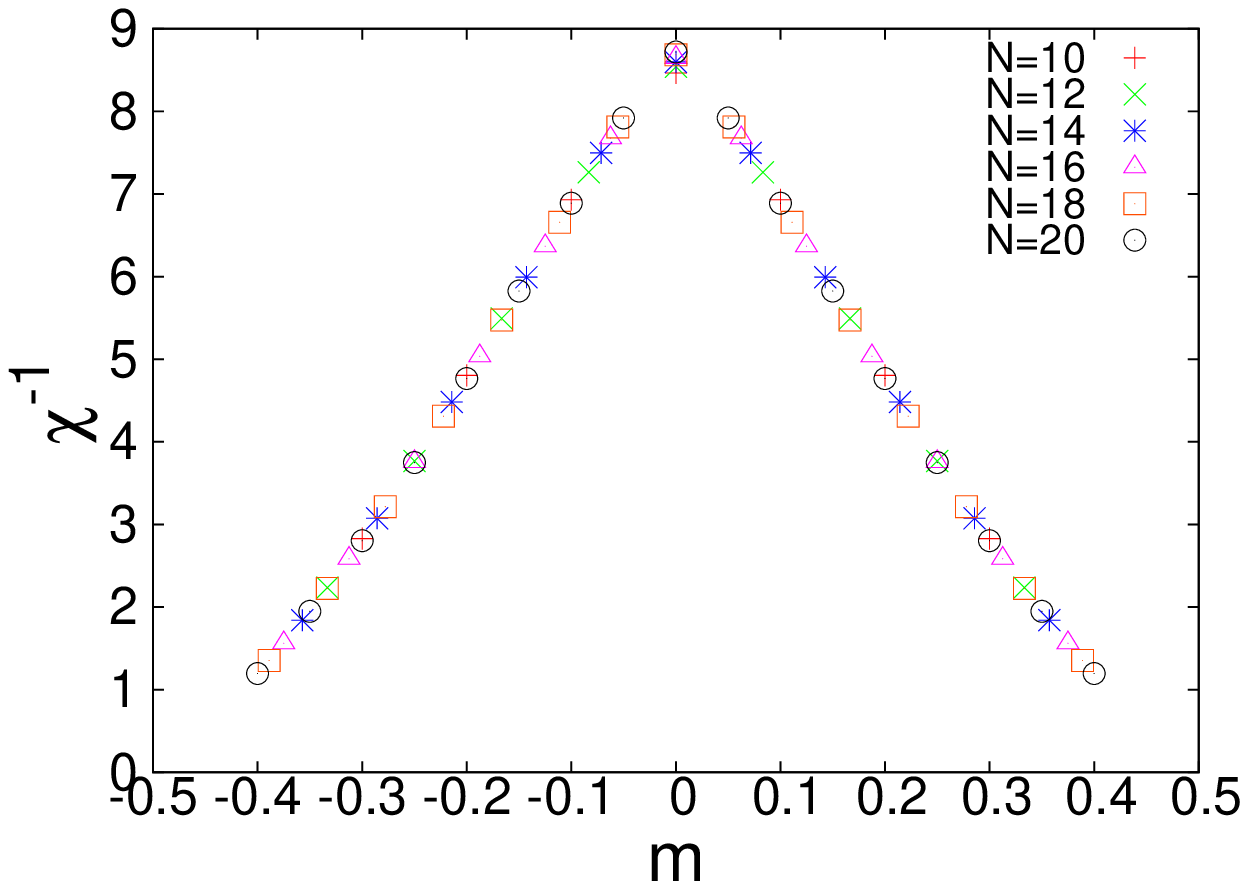}
  \subcaption{$\Delta=1$}
  \label{zikagyaku1}
\end{minipage}
\begin{minipage}[t]{0.32\linewidth}
 \centering
  \includegraphics[keepaspectratio,scale=0.48]{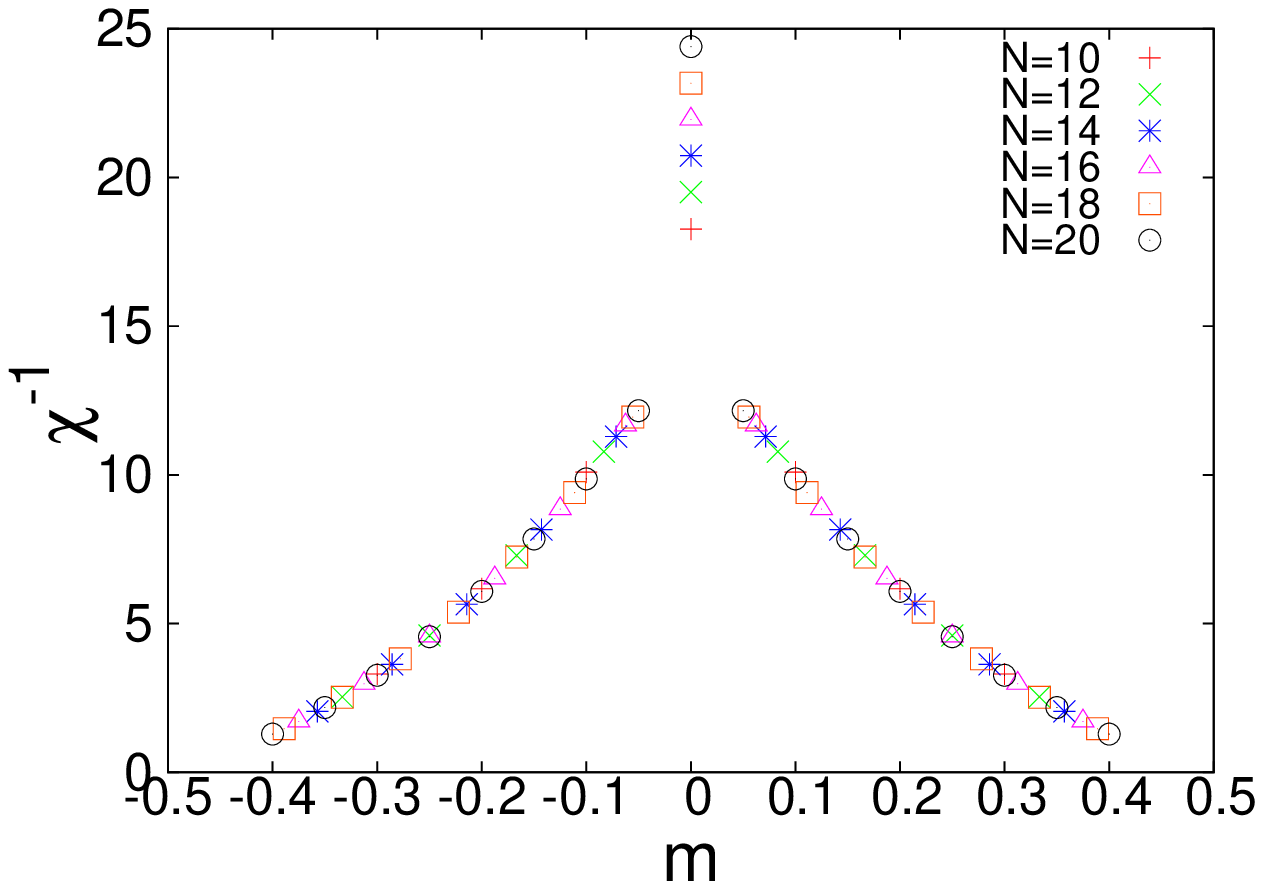}
 \subcaption{$\Delta=2$}
 \label{zikagyaku2}
\end{minipage}
\caption{
Magnetization dependence of the inverse of the magnetic susceptibility $\chi^{-1}$
of the $S=1/2$ XXZ antiferromagnetic chain for several system sizes $N$: 10, 12, 14, 16, 18, and 20.
\subref{zikagyaku1} demonstrates that $\chi^{-1}$ shows a cusp and
a low size dependence at zero magnetization.
\subref{zikagyaku2} demonstrates that $\chi^{-1}$ shows a sharp cusp of
large positive values and high size dependence at zero magnetization,
compared with \subref{zikagyaku1}.
Therefore, it is easier to observe the cusp of $\chi^{-1}$ than that of $\chi$ in Fig.~\ref{chi}.
Moreover, it is possible for $\chi^{-1}$ to show an anomaly for $\Delta>1$
owing to its high size dependence at zero magnetization.
}
\label{chiin}
\end{figure*}

\section{Numerical Results}
We calculate the lowest-energy eigenvalue $E(N,M)$ to derive
the magnetic susceptibility $\chi$ and the fourth derivative $A$,
using numerical diagonalization by TITPACK Ver.2\cite{37}
and $H\phi$.\cite{35}
As an example, we treat an $S=1/2$ XXZ antiferromagnetic spin chain 
\begin{align}
 \mathscr{\hat{H}} &= J\sum_{j=1}^{N} (\hat{S}_{j}^{x}\hat{S}_{j+1}^{x}+
 \hat{S}_{j}^{y}\hat{S}_{j+1}^{y}+ \Delta\hat{S}_{j}^{z}\hat{S}_{j+1}^{z}),
\label{1}
\end{align}
where $\hat{S}_j^x,\hat{S}_j^y,\hat{S}_j^z$ is the
$j$th site spin operator in the $x,y,z$ direction.
$\Delta$ is an anisotropic parameter that
takes a 0.1 increment of values from 0 to 2.
The phase of the chain is changed by $\Delta$, which shows ferromagnetic phase for $\Delta \le-1$, Tommonaga--Luttinger phase for $-1 < \Delta \le 1$, and antiferromagnetic phase for $\Delta>1$.
$N$ is even from 10 to 26.
We then give an exchange interaction $J=1$.
The boundary condition of the model is periodic:
\begin{align}
 \hat{S}_{N+1} = \hat{S}_{1}.
\end{align}
In this section, we present our numerical data for $\Delta=0,1,2$
with several sizes from 10 to 20.
\begin{figure*}[t]
\begin{minipage}[t]{0.33\linewidth}
 \centering
  \includegraphics[keepaspectratio,scale=0.48]{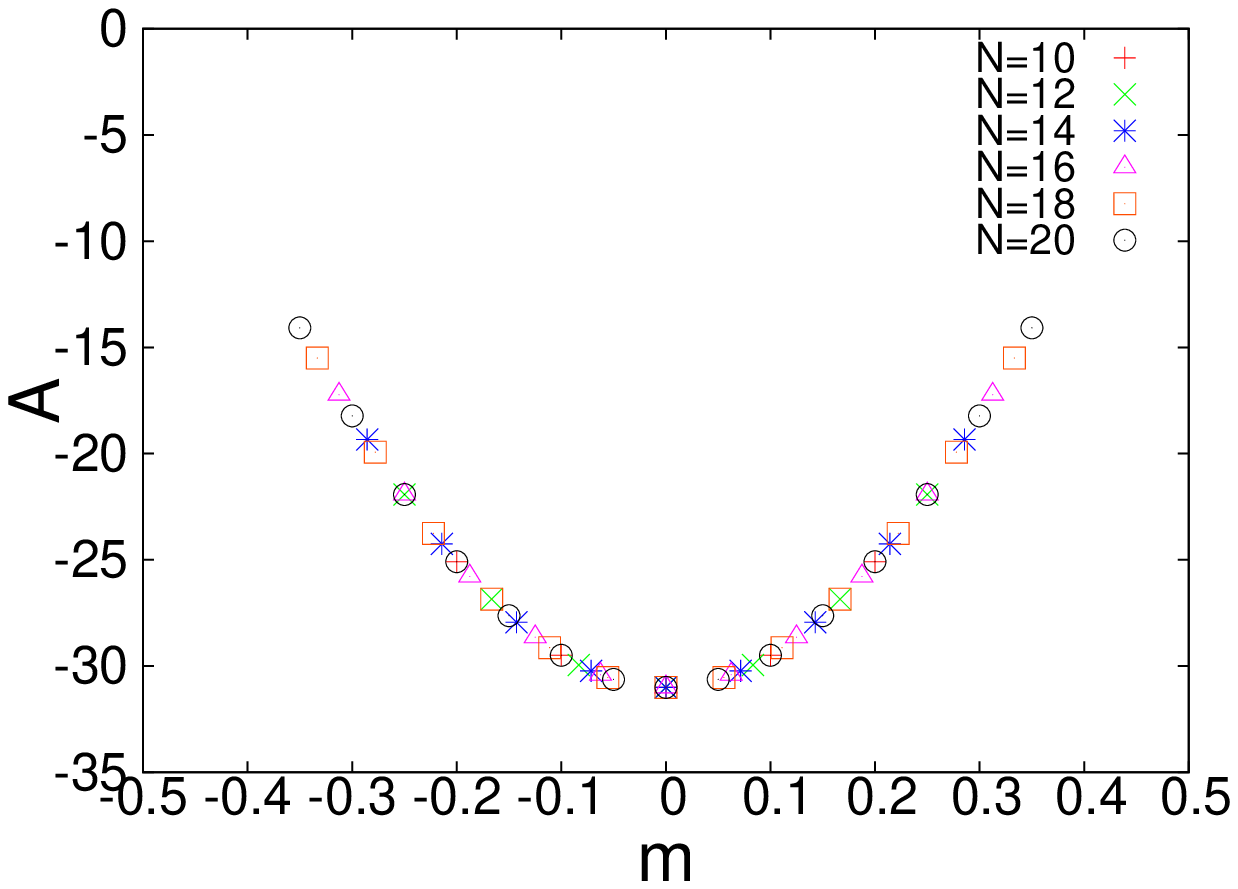}
 \subcaption{$\Delta=0$}
 \label{A0}
\end{minipage}
\begin{minipage}[t]{0.33\linewidth}
 \centering
    \includegraphics[keepaspectratio,scale=0.48]{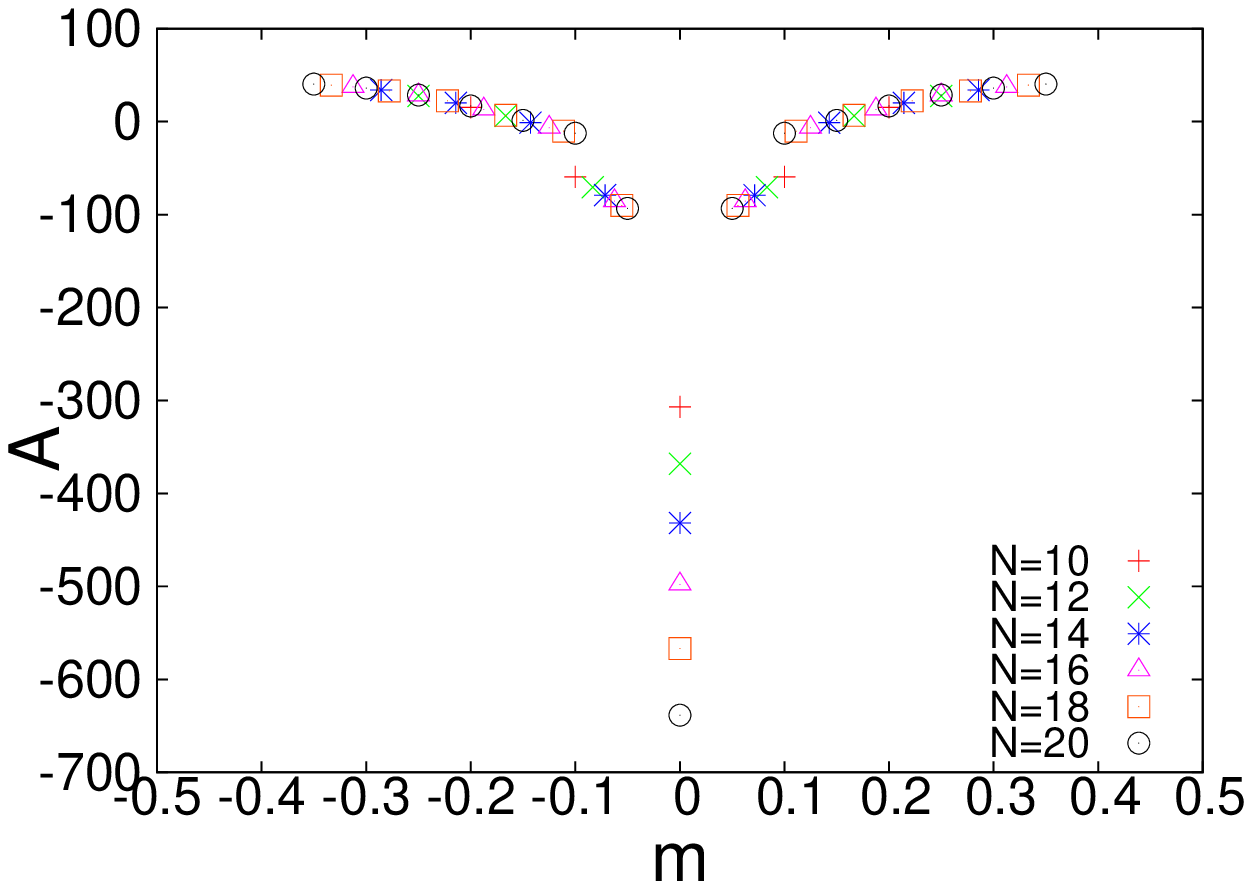}
 \subcaption{$\Delta=1$}
 \label{A1}
\end{minipage}
\begin{minipage}[t]{0.32\linewidth}
 \centering
    \includegraphics[keepaspectratio,scale=0.48]{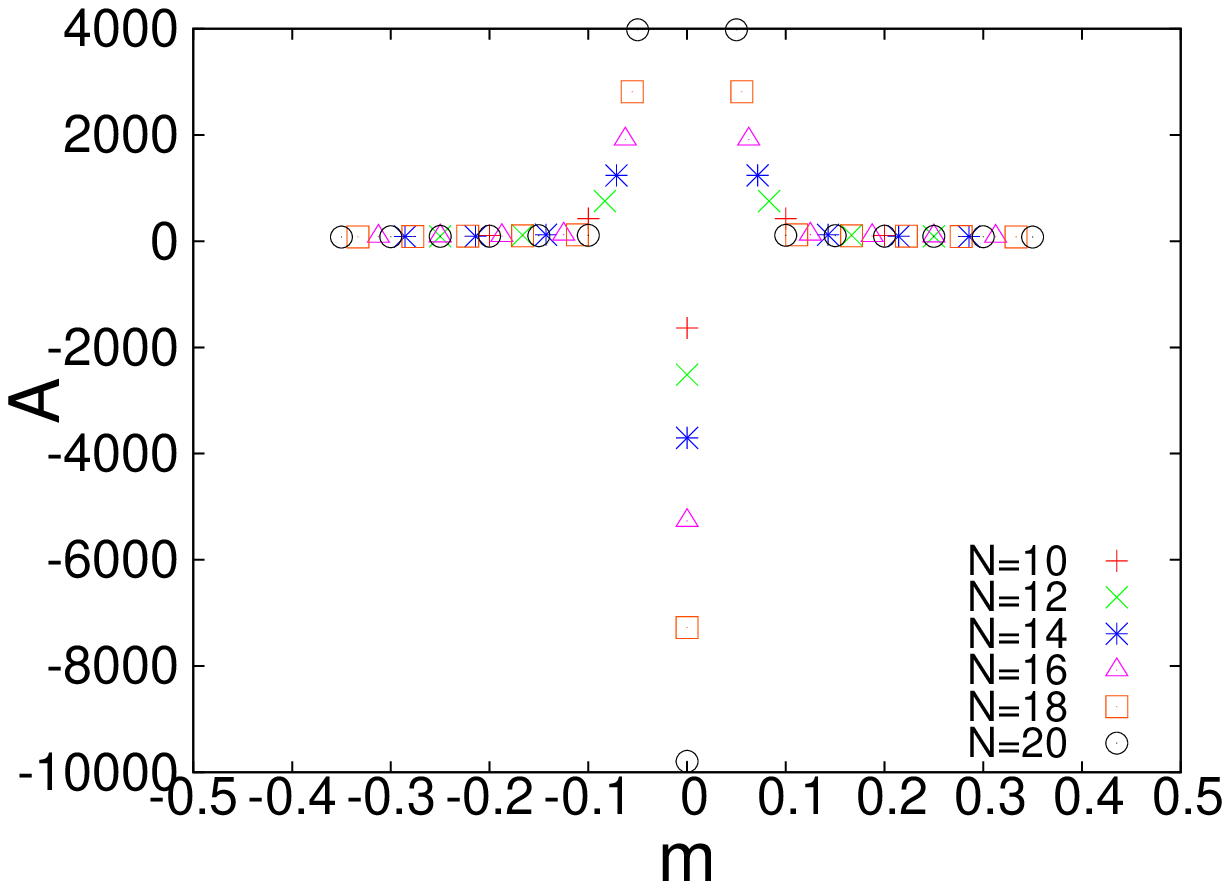}
 \subcaption{$\Delta=2$}
 \label{A2}
\end{minipage}
\caption{
Magnetization dependence of the fourth derivative $A$ of the $S=1/2$ XXZ
antiferromagnetic chain for several system sizes $N$: 10, 12, 14, 16, 18, and 20.
Both \subref{A1} and \subref{A2} demonstrate that $A$ shows a sharp cusp of
large negative values and a high size dependence at zero magnetization.
Furthermore, \subref{A2} shows that $A$ has large positive values
as the magnetization approaches zero.
Therefore, it is possible for $A$ to show an anomaly for $\Delta\ge1$
owing to its high size dependence at zero magnetization.
}
\label{A}
\end{figure*}

\subsection{Magnetic susceptibility\ $\chi$ and $\chi^{-1}$}
First, we show the magnetization dependence of $\chi$ in Fig.~\ref{chi}.
Figures~\ref{chi}\subref{zika0} and \ref{chi}\subref{zika1} show smooth curves.
Figure~\ref{chi}\subref{zika2} only shows a sharp cusp at zero magnetization.
However, this cusp does not indicate an anomaly, as
an anomaly must satisfy the following conditions:
(1) $\chi$, $\chi^{-1}$, and $A$ have a cusp, and (2) the size dependence of
the cusp is large in the thermodynamic limit.
Thus, Fig.~\ref{chi}\subref{zika2} does not show the anomaly
as the size dependence is small.
Similarly, neither Fig.~\ref{chi}\subref{zika0} nor Fig.~\ref{chi}\subref{zika1}
shows the anomaly.
The results demonstrate that the anomaly of $\chi$ is not shown.

Next, the magnetization dependence of $\chi^{-1}$ is shown in Fig.~\ref{chiin}.
Figure~\ref{chiin}\subref{zikagyaku0} does not show an anomaly because
the graph does not have a cusp.
Figures~\ref{chiin}\subref{zikagyaku1} and \ref{chiin}\subref{zikagyaku2}
have sharp cusps at zero magnetization, compared with Fig.~\ref{chi}.
Thus, it is clearer to observe the cusp of $\chi^{-1}$ than of $\chi$.
However,
Fig.~\ref{chiin}\subref{zikagyaku1} does not show an anomaly as the
size dependence is small at zero magnetization.
In contrast, Fig.~\ref{chiin}\subref{zikagyaku2}
demonstrates the possibility of showing an anomaly because the size dependence is large. 
The results indicate a possibility that $\chi^{-1}$ shows an anomaly
for $\Delta>1$ in the thermodynamic limit.
The details are discussed in a later section.

\begin{figure}[t]
 \begin{center}
  \includegraphics[keepaspectratio,scale=0.45]{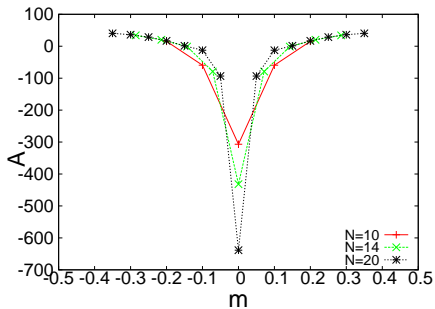}
 \end{center}
 \caption{Magnetization dependence of the fourth derivative
 $A$ for $\Delta=1$ for three system sizes: 10, 14, and 20.
 The $A$ values form a continuous curve, in contrast with Fig.~\ref{A}\subref{A1}.
 }
 \label{4DN21sys3}
\end{figure}

\begin{figure*}[t]
 \begin{center}
  \includegraphics[keepaspectratio,scale=0.45]{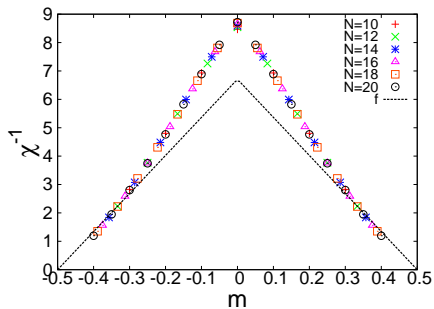}
 \end{center}
 \caption{Magnetization dependence of the inverse of the magnetic susceptibility
 $\chi^{-1}$ for $\Delta=1$.
 $f$ is the fitting function expressed by Eq.~(\ref{11}).
 Our numerical data are consistent with $f$ near saturation magnetization.}
 \label{zikagyaku1ta}
\end{figure*}

\subsection{Fourth derivative $A$}
We show the magnetization dependence of $A$ in Fig.~\ref{A}, which indicates
that $A$ decreases as the magnetization approaches zero for $0\le\Delta\le1$.
Figure~\ref{A}\subref{A0} does not show an anomaly as the graph does not have a cusp.
Figures~\ref{A}\subref{A1} and \ref{A}\subref{A2} have sharp cusps
at zero magnetization in comparison to Fig.~\ref{chiin}.
Furthermore, these graphs show the possibility that $A$ at zero magnetization
indicates an anomaly as its size dependence is large.
This shows that it is easier to observe the possibility that an anomaly
exists for $A$ than for $\chi^{-1}$.
The difference between the graphs is the behavior of $A$ at $m=\pm 1/N$.
Figure~\ref{A}\subref{A1} demonstrates that at $m=\pm 1/N$, $A$ exhibits negative values.
Although $A$ in Fig.~\ref{A}\subref{A1} appears to be discontinuous
near $m=0.1$, this behavior is superficial.
In fact, Fig.~\ref{4DN21sys3} shows that near $m=0.1$, $A$ is continuous
for three system sizes: 10, 14, and 20.
Thus, $A$ near $m=0.1$ is continuous for $\Delta=1$.
In contrast, Fig.~\ref{A}\subref{A2} demonstrates that $A$ at $m=\pm 1/N$
shows large positive values, in contrast to the large negative value of $A$ at zero magnetization.
The behavior of $A$ indicates the possibility of showing an anomaly as its size dependence is large.
This is explained by Eq.~(\ref{eq:yonkaigap2}).
However, we do not understand how the behavior of $A$ for $\Delta<1$ changes in the thermodynamic limit.
These details are discussed in a later section.

\section{Comparison with exact solutions}
In this section, we compare our numerical data with exact solutions\cite{9,14,15,16,24}
to investigate the behavior of $A$.
The behavior of $\chi$ is well established for all $\Delta$.
However, the behavior of $A$ has not previously been studied.
The reliability of the data of $A$ increases when the data of $\chi$ agree with
exact solutions.
This leads to investigation of the behavior of $A$.

\subsection{Comparison with magnetic susceptibility
near saturation magnetization}
For the spin $S$ antiferromagnetic chain,
the inverse of magnetic susceptibility $\chi^{-1}$ is proportional to a magnetization $S-m$
near the saturation magnetization.\cite{2} In our case, it is shown that
\begin{align}
\label{11}
 \chi^{-1} \propto 1/2-m.
\end{align}
We investigate whether our numerical data are consistent with Eq.~(\ref{11}).
Figure~\ref{zikagyaku1ta} shows a magnetization dependence of $\chi^{-1}$
for $\Delta=1$ with a fitting function $f$ that is described by Eq.~(\ref{11}).
The fitting is performed under $0.3\le |m| \le0.4$.
The graph demonstrates a linear relation between $\chi^{-1}$ and $m$
near $m=1/2$ because $f$ is consistent with our data.
However, the relation is not applied to the points where $f$ is not
consistent with our data.
Thus, our data of $\chi$ are reliable, and the reliability of our data of $A$
increases.

\begin{figure*}[t]
\begin{minipage}[b]{0.45\linewidth}
 \centering
  \includegraphics[keepaspectratio,scale=0.5]{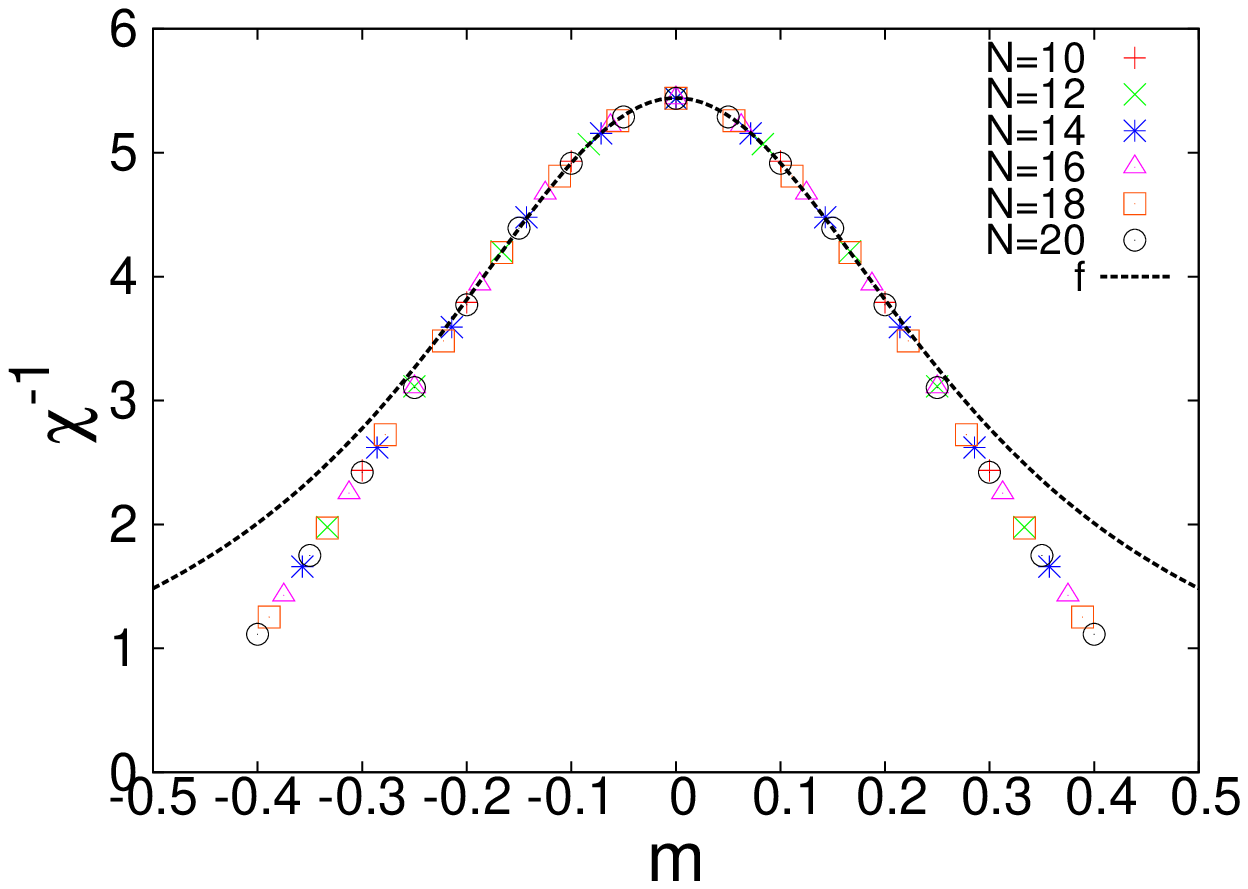}
 \subcaption{$\Delta=1/2$}
 \label{chiin05}
\end{minipage}
\begin{minipage}[b]{0.45\linewidth}
 \centering
  \includegraphics[keepaspectratio,scale=0.5]{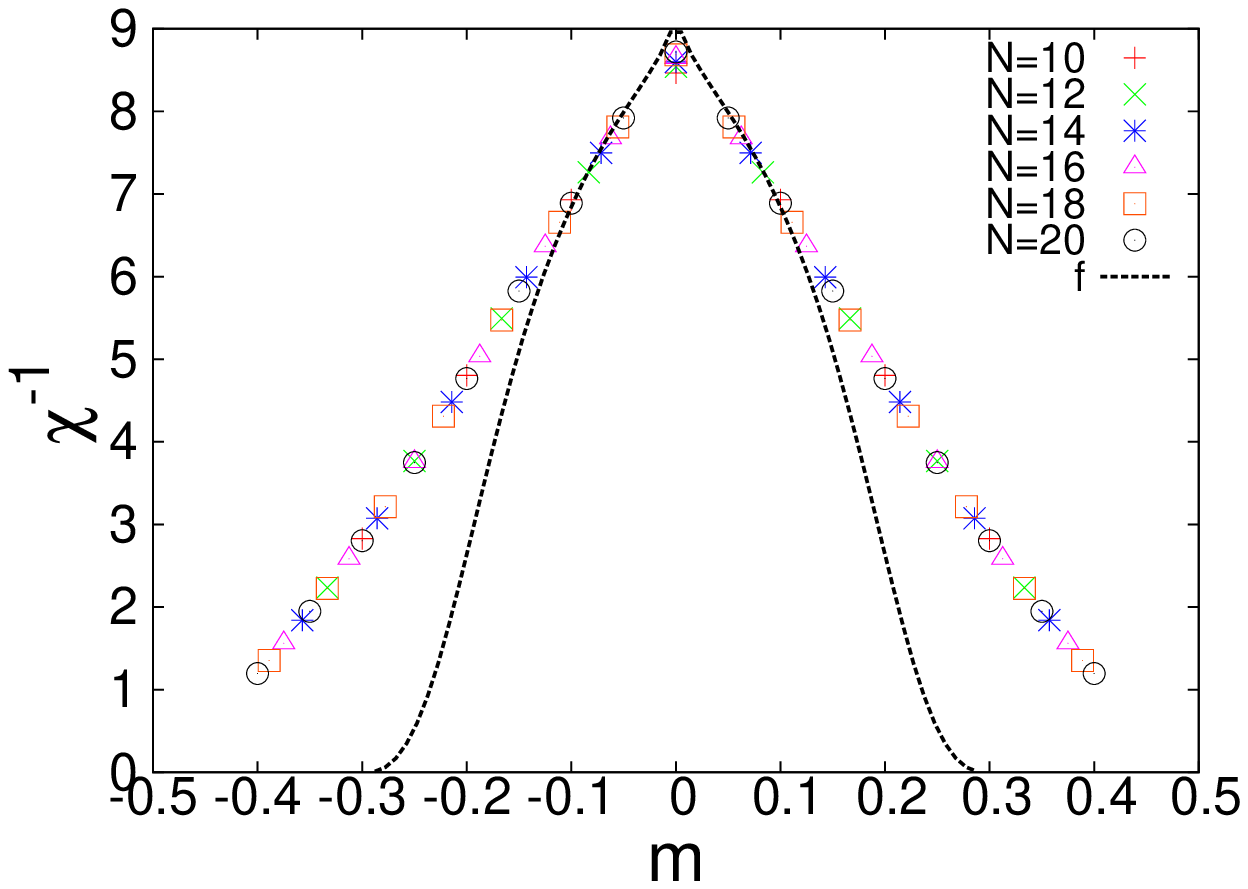}
  \subcaption{$\Delta=1$}
  \label{chiin1}
\end{minipage}
\caption{
Magnetization dependence of the inverse of the magnetic susceptibility
$\chi^{-1}$ with a fitting function $f$.
\subref{chiin05} $f$ is expressed by the inverse of Eq.~(\ref{14}). 
\subref{chiin1} $f$ is expressed by the inverse of Eq.~(\ref{hubbardv2}).
Our numerical data are consistent with $f$ under $0 < |m| \le0.1$.
}
\label{chiinfit}
\end{figure*}

\begin{figure*}[t]
\begin{minipage}[b]{0.45\linewidth}
 \centering
  \includegraphics[keepaspectratio,scale=0.5]{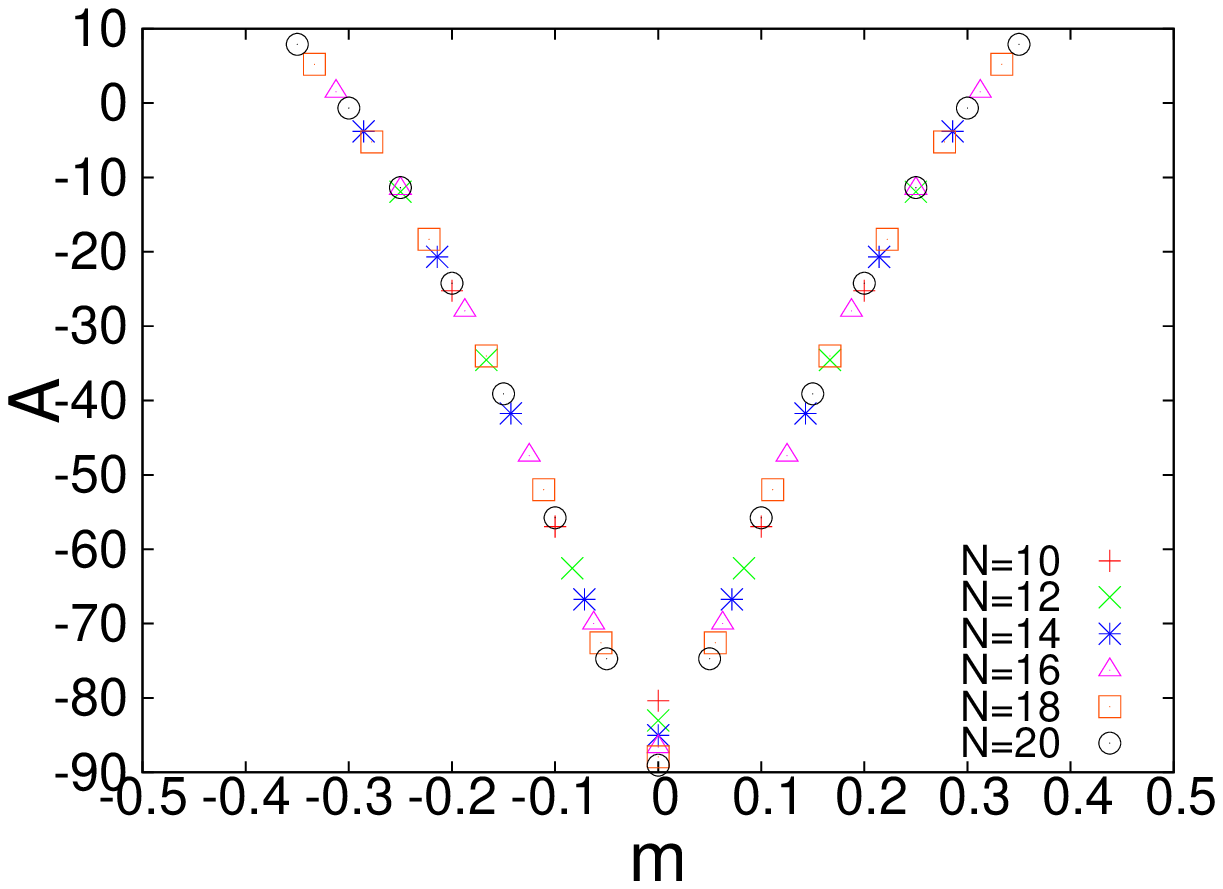}
 \subcaption{$\Delta=0.4$}\label{A04}
\end{minipage}
\begin{minipage}[b]{0.45\linewidth}
 \centering
  \includegraphics[keepaspectratio,scale=0.5]{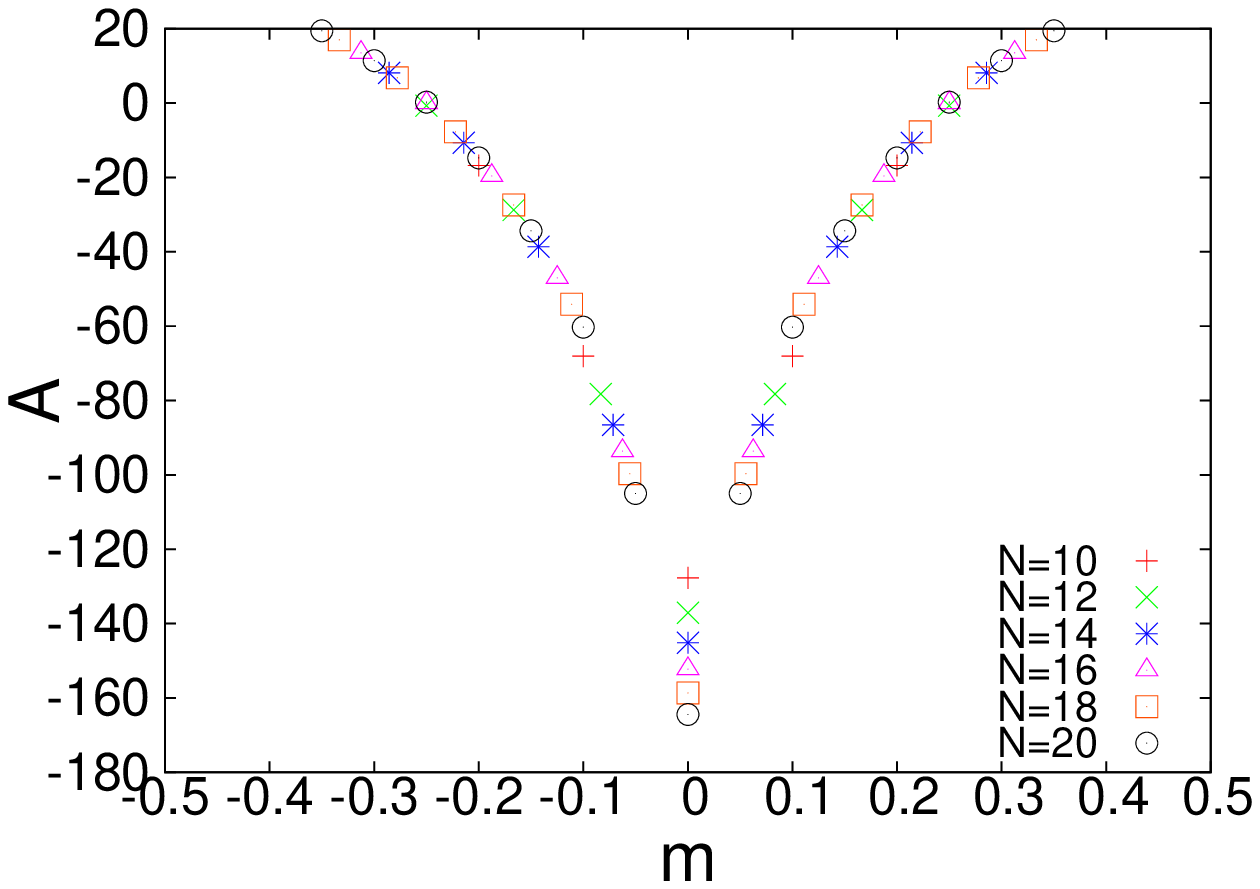}
  \subcaption{$\Delta=0.6$}\label{A06}
\end{minipage}
\caption{
Magnetization dependence of the fourth derivative $A$ near $\Delta$$=$$1/2$.
Our numerical data are consistent with Eq.~(\ref{n2fi}) and Eq.~(\ref{n2inf}).
}
\label{Aexact}
\end{figure*}

\subsection{Comparison with Bethe-ansatz solution}
The Bethe ansatz is an exact method applied in a wide range of fields, such as
quantum field theory and statistical mechanics.
We compare our numerical data with exact solutions.
The Zeeman energy is given by
\begin{align}
\label{zeeman}
\mathscr{\hat{H}}_{z} &= h\sum_{j=1}^{N}\hat{S}_{j}^{z}.
\end{align}
$\mathscr{\hat{H}}_{z}$ and the Hamiltonian in Eq.~(\ref{1}) commute.
$h$ is a magnetic field.
\subsubsection{Case of $\chi^{-1}$ for $0<$$\Delta$$\le 1$}
First, under $0<\Delta<1$, the exact solution of $\chi$ is given by\cite{22}
\begin{align}
 \chi &= \F{4\gamma}{\pi(\pi-\gamma)\sin \gamma}\{1+ 
 \mathcal{O}(h^2) + \mathcal{O}(h^{\F{4\gamma}{\pi-\gamma}})\},\label{12}\\
 \gamma &= \arccos\Delta \label{13}.
\end{align}
We then rewrite Eq.~(\ref{12}) as a function of $m$
as it is difficult to compare our numerical data with Eq.~(\ref{12}):
\begin{align}
  \chi &= \F{4\gamma}{\pi(\pi-\gamma)\sin \gamma}
  +c_{1}m^2+c_{2}|m|^{\F{4\gamma}{\pi - \gamma}},\label{14}\\
  h &= \F{4\gamma}{\pi(\pi-\gamma)\sin \gamma}m,\nonumber
  \end{align}
where $c_{1}$ and $c_{2}$ are constants.
We perform the fitting with Eq.~(\ref{14}) under $0< |m| \le0.1$.
The result of this fitting is shown in Fig.~\ref{chiinfit}\subref{chiin05}.
Figure~\ref{chiinfit}\subref{chiin05} indicates that our data are consistent
with exact solutions near zero magnetization.
Therefore, this consistency increases the reliability of our numerical data of $\chi$
for $0<\Delta<1$.

Next, we explain the exact solution of $\chi$ for $\Delta=1$.
In this case, although it appears that $h^{\F{4\gamma}{\pi - \gamma}}=h^0$ from Eq.~(\ref{12}), there remains a possibility of logarithmic behavior from the solutions of a Hubbard model.\cite{6}
The Hubbard model is regarded as an isotropic Heisenberg model
with an infinite Coulomb repulsion.
Thus, using the exact solution of the Hubbard model,
that of $\chi$ is given by\cite{6}
\begin{align}
\label{hubbard}
 \F{\chi}{\chi_{0}} &= 1 + \F{1}{2}\F{1}{\ln{\F{h_{c}}{h}\gamma_{c}}} 
 - \F{1}{4}\F{\ln{\ln{\F{h_{c}}{h}\gamma_{c}}}}{(\ln{\F{h_{c}}{h}\gamma_{c}})^2}
 + \mathrm{h.o.},\\
 \label{hcgc}
 h_{c} &= 4\sin^2\left(\F{\pi}{2}n\right),
 \ \gamma_{c}=\F{\pi}{2}\sqrt{\F{2\pi}{e}},
\end{align}
where $\chi_{0}$, $n$, and h.o. are the magnetic susceptibility
at zero magnetization, a filling that denotes electron density,
and high-order terms, respectively.
For an isotropic Heisenberg model, $n$$=$$1$ and $\chi_{0}$$=$$1/\pi^2$.\cite{4,7}
Similarly, we rewrite Eq.~(\ref{hubbard}) as a function of $m$
for the Heisenberg model
\begin{align}
\label{hubbardv2}
 \chi&= \F{1}{\pi^2} + \F{1}{2\pi^2}\F{1}{\ln{\F{2\pi\sqrt{\F{2\pi}{e}}}{d_{0}m}}} 
 - \F{1}{4\pi^2}\F{\ln{\ln{\F{2\pi\sqrt{\F{2\pi}{e}}}{d_{0}m}}}}
 {\left(\ln{\F{2\pi\sqrt{\F{2\pi}{e}}}{d_{0}m}}\right)^2},\\
 h &=d_{0}m, \nonumber
\end{align}
where $d_{0}$ is a constant.
We perform the fitting with Eq.~(\ref{hubbardv2}) under
$0\le |m| \le0.1$.
The result is shown in Fig.~\ref{chiinfit}\subref{chiin1}.
Figure~\ref{chiinfit}\subref{chiin1} indicates that our data are
consistent with exact solutions near zero magnetization.
Thus, our data are consistent with exact solutions in $0<\Delta\le1$.
This indicates that the reliability of the data of $A$ increases with that of $\chi$.
However, for $\Delta=1$ the magnetic susceptibility shows an infinite slope when $m$ approaches zero from Griffiths's theory. The cause of the differences between the theoretical and calculated results is a finite size effect.

\subsubsection{Case of $A$}
The exact solutions of $A$ have not been investigated.
However, C.N. Yang and C.P. Yang discussed\cite{5,31}
\begin{empheq}[left={\displaystyle \lim_{m \to 0+}A=
\empheqlbrace}]
{alignat=2}
\mbox{finite} & \quad (-1<\Delta<1/2) \label{n2fi}\\
\mbox{infinite} & \quad (1/2<\Delta<1) \label{n2inf}.
\end{empheq}
The numerical data of $A$ are shown in Fig.~\ref{Aexact}.
Figure~\ref{Aexact}\subref{A04} indicates that $A$ becomes finite
as $m$ approaches zero because its size dependence is small.
In contrast, Fig.~\ref{Aexact}\subref{A06} indicates that $A$ appears to
become infinite as $m$ approaches zero from
of its large size dependence.
Thus, our data are explained by the tendency of the exact solutions.
$\Delta$$=$$1/2$ corresponds to $N=2$ supersymmetry (SUSY) in
a conformal field theory.\cite{36}
Details of this are discussed in a later section. 

\section{Anomaly and Correction Term associated with Conformal Field Theory}
\begin{figure*}[ht]
 \begin{center}
  \includegraphics[keepaspectratio,scale=0.5]{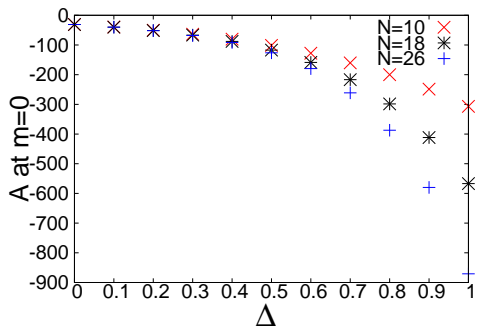}
 \end{center}
 \caption{Anisotropic parameter $\Delta$ dependence of the fourth derivative $A$
 at zero magnetization for several system sizes $N$: 10, 18, and 26.
 $A$ appears to diverge for $\Delta>1/2$ as the size dependence is large.
 Thus, the starting point of the anomaly of $A$ is the $\Delta=1/2$ point.
 The anomaly of $A$ at $1/2<\Delta<1$ does not indicate the phase transition from the scaling dimension.\cite{45} The $-1<\Delta<1/2$ region is named TL phase (I) and $1/2<\Delta\le1$ is named TL phase (I\hspace{-.1em}I) based on the behavior of $A$.}
 \label{4D0D}
\end{figure*}

In this section, we describe the relationship between the anomaly of $A$ and
a conformal field theory (CFT).
In addition, the correction term obtained from the CFT is discussed.
\subsection{Anomaly at $\Delta=1/2$}
We demonstrate that the $\Delta=1/2$ point corresponds to the $N=2$ SUSY in the CFT.
We apply the CFT to the $S=1/2$ XXZ chain.
The anisotropic parameter of the chain $\Delta$ is related to the scaling dimension
$x_T$, which is associated with the critical exponent.\cite{43}
It is shown for $-1<\Delta\le1$ that\cite{42,44}
\begin{align}
	x_{T}(k=0) &= \F{2\pi}{\arccos{(-\Delta)}}, \label{scd}
\end{align}
where $k$ is the wavenumber of the spin state that is a parameter obtained from translational symmetry.
We focus on the scaling dimension with the wavenumber $k=0$ and zero magnetization, as it is compatible with the symmetry of the Hamiltonian.
For $\Delta=1/2$, from Eq.~(\ref{scd}), $x_T$($k=0$)$=$3.
The scaling dimensions $x_{T}(k=0) > 2$ have irrelevant characteristics.\cite{43} 
Thus, $x_{T}(k=0)=3$ shows the irrelevant characteristics.
S.K. Yang\cite{40} demonstrated that $x_{T}(k=0)=3$ corresponds to $N=2$ SUSY
from the correspondence between the XXZ chain and Ashkin--Teller model.
Later, P. Ginsparg\cite{36} showed the same correspondence 
from the relation between the XXZ chain and free boson model.
Therefore, these discussions show that $\Delta=1/2$ corresponds to $N=2$ SUSY.

\subsection{Anomaly and scaling dimensions}
We show that the anomaly of $A$ is influenced by scaling dimensions in the CFT.
In the CFT, the energy gap $\Delta E$ for a finite system size $N$ is given by\cite{46}
\begin{align}
 \Delta E = \F{2 \pi v}{N} \left\{ x + C_1 \left(\F{1}{N} \right)^{x_{T} -2}
  + C_2 \left(\F{1}{N} \right)^{2({x_{T} -2})} \right\}, \label{egcft}
\end{align}
where $x$ is a scaling dimension that is different from $x_T$, $C_1$ and $C_2$
are constants, and $v$ is the velocity of the spin wave.
For a sine-Gordon model that corresponds to the XXZ chain, $C_1 =0$, $x_T>2$,\cite{39} and $C_2<0$.\cite{42}
First, we consider $\epsilon^{(4)}(N, m)$ at $x_T>2$.
Next, considering $1/N \propto m$, Eq.~(\ref{egcft}) is written as
\begin{align}
 \Delta E = \F{2 \pi v}{N} \left\{ x + C_2 m^{2({x_{T} -2})} \right\}. \label{egcft2}
\end{align}
Substituting Eq.~(\ref{egcft2}) into Eq.~(\ref{nikai}), the second differentials are written by
\begin{align}
 \epsilon^{(2)}(N, m) = 4\pi v \left\{ x + C_2 |m|^{2({x_{T} -2})} \right\}, \label{nikaicft2}
\end{align}
where $m$ is replaced by $|m|$ as the system has the spin reversal symmetry $m\rightarrow -m$.
This equation is consistent with the exact solution of Eq.~(\ref{14}).
The fourth differential of magnetization with respect to energy is then written in the form 
\begin{align}
 \epsilon^{(4)}(N, m) = 4\pi v C_2 (2x_{T}-4)(2x_{T}-5) |m|^{2({x_{T} -3})}. \label{yonkaicft2}
\end{align}
$\epsilon^{(4)}(N, m)$ diverges for $2<x_{T}<3$, i.e., $1/2<\Delta<1$ in the thermodynamic limit.
$\epsilon^{(4)}(N, m)=0$ at $x_T=5/2$ is a subject of future works as few investigations have been carried out.
In addition, we focus on the size dependence of $\epsilon^{(4)}(N, m=0)$. Considering the Gaussian model,\cite{42,47} we extend Eq.~(\ref{egcft2}) in the form 
\begin{align}
 &E(N, M) - 2E(N, 0) +E(N, -M)\nonumber\\
 &=E(N, M) - E(N, 0) -(E(N, -M) -E(N, 0))\nonumber \\
 &=\F{4\pi v |M|^2}{N}\left\{x + C_2 \left|\F{M}{N} \right|^{2(x_T -2)} \right\}, \label{nikaiex}
\end{align}
where $|M|$$\ll$$N$.
Although few investigations into the extension of Eq.~(\ref{egcft2}) have been carried out, with respect to $|M/N|^{2(x_T -2)}$, we consider that the average distance between quasi-particles is $|N/M|$, which is renormalized by $x_T$.
The relation is applied for the fourth differential in Eq.(\ref{yonkai}) in the form 
\begin{align}
 &\epsilon^{(4)}(N, m=0)\nonumber\\
 &= N^3 \{E(N, 2) - 2E(N, 0) +E(N, -2) \nonumber\\
 &-4(E(N, 1) - 2E(N, 0) +E(N, -1))\}\nonumber\\
 &=16\pi v C_2 (2^{2(x_T -2)}-1)N^{6-2x_T}, \label{yonkaiex}
\end{align}
where the coefficient $16\pi v C_2 (2^{2(x_T -2)}-1)$ has a negative value for $C_2<0$ and $x_T>2$.
Thus, $\epsilon^{(4)}(N, m=0)$ diverges for $2<x_T<3$ in the thermodynamic limit.

Next, we explain $\epsilon^{(4)}(N, m)$ at $x_T=2$, i.e., $\Delta=1$.
The energy gap $\Delta E$ is written as\cite{16}
\begin{align}
  \Delta E = \frac{2 \pi v}{N}\F{1}{2}
  \left(1-\frac{1}{2}\frac{1}{\ln{\F{N}{N_0}}} + \frac{1}{4}\frac{\ln\left(\ln{\F{N}{N_0}}\right)}{\left(\ln{\F{N}{N_0}}\right)^{2}}  \right), \label{eq:xt2E}
\end{align}
where $N_0$ is a non-universal renormalization constant.
Under $1/N\propto m$, substituting Eq.~(\ref{eq:xt2E}) into Eq.~(\ref{nikai}), we obtain
\begin{align}
  \epsilon^{(2)}(N, m)
  = 2\pi v \left( 1-\frac{1}{2}\frac{1}{\ln{\F{m_0}{m}}}  +\frac{1}{4}\frac{\ln\left(\ln{\F{m_0}{m}}\right)}{\left(\ln{\F{m_0}{m}}\right)^{2}}\right),
  \label{eq:nikaixt2}
\end{align}
where $m_0$ is a constant that corresponds to $1/N_0$.
Hence, the fourth differentials are shown by
\begin{align}
\epsilon^{(4)}(N, m)&=\pi v \left(m \ln{\F{m_0}{m}}\right)^{-2}\left(1-  
 \frac{\ln(\ln{\F{m_0}{m}}))+3/2}{ \ln{\F{m_0}{m}}}\right. \nonumber\\
&\left. +\mathcal{O}\left(\left(\ln{\F{m_0}{m}}\right)^{-2}\right)  \right).
                \label{eq:yonkaixt2mnon0}
\end{align}
Thus, $\epsilon^{(4)}(N, m)$ diverges in the thermodynamic limit.
Furthermore, we discuss the size dependence of $\epsilon^{(4)}(N,m=0)$.
We use Eq.~(\ref{eq:xt2E}) in the same procedure as for Eq.~(\ref{nikaiex}) and obtain
\begin{align}
  &E(N, M) -2 E(N, 0)+E(N, -M) \nonumber\\
  &=\frac{2 \pi v}{N} |M|^{2} \left(1 -\F{1}{2}\F{1}{\ln{\F{N}{N_0|M|}}}
  + \F{1}{4}\F{\ln(\ln{\F{N}{N_0|M|}})}{(\ln{\F{N}{N_0|M|}})^2} \right).
 \label{eq:energygapxt2}
\end{align}
Using this relation, the fourth differentials are expressed by
\begin{align}
  &\epsilon^{(4)} (N, m=0) \nonumber \\
  &= -4 \pi v N^{2} \left(\ln{\F{N}{N_0}}\right)^{-2} \ln{2} \nonumber \\
  &\times \left(1 - \frac{\ln(\ln{\F{N}{N_0}})-\ln{2} -1/2}{\ln{\F{N}{N_0}}}
  + \mathcal{O}\left(\left(\ln{\F{N}{N_0}}\right)^{-2} \right)\right). \label{eq:yonkaixt2}
\end{align}
Therefore, $\epsilon^{(4)}(N, m=0)$ diverges in the thermodynamic limit.

These facts indicate that the scaling dimension $x_T$ influences the energy gap, magnetic susceptibility, and fourth derivative for a finite magnetization. The anomalies of $\chi$ and $A$ are subject to the change in scaling dimension that is related to phase transition.

\begin{figure*}[ht]
\begin{minipage}[t]{0.33\linewidth}
 \centering
  \includegraphics[keepaspectratio,scale=0.48]{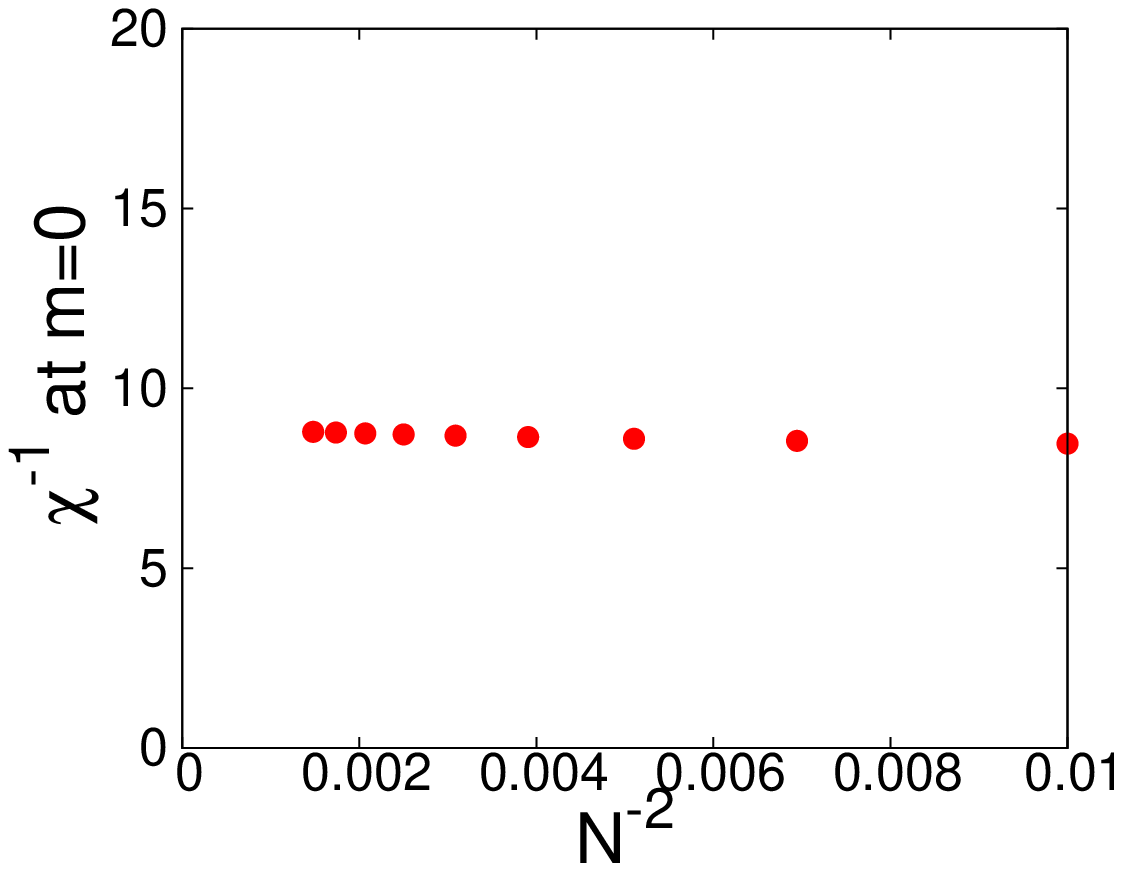}
 \subcaption{$\Delta=1$}
 \label{chiinN21}
\end{minipage}
\begin{minipage}[t]{0.33\linewidth}
 \centering
  \includegraphics[keepaspectratio,scale=0.48]{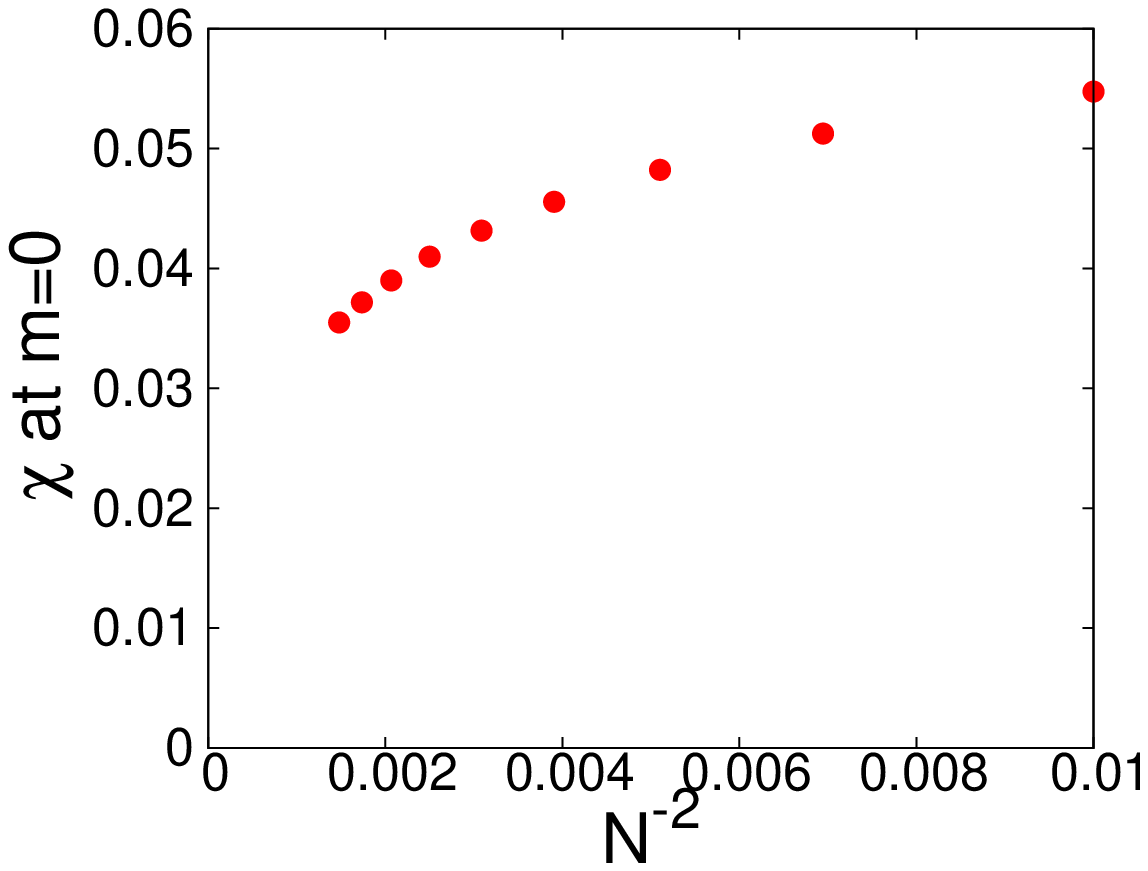}
  \subcaption{$\Delta=2$}
  \label{chiN22}
\end{minipage}
\begin{minipage}[t]{0.32\linewidth}
 \centering
  \includegraphics[keepaspectratio,scale=0.48]{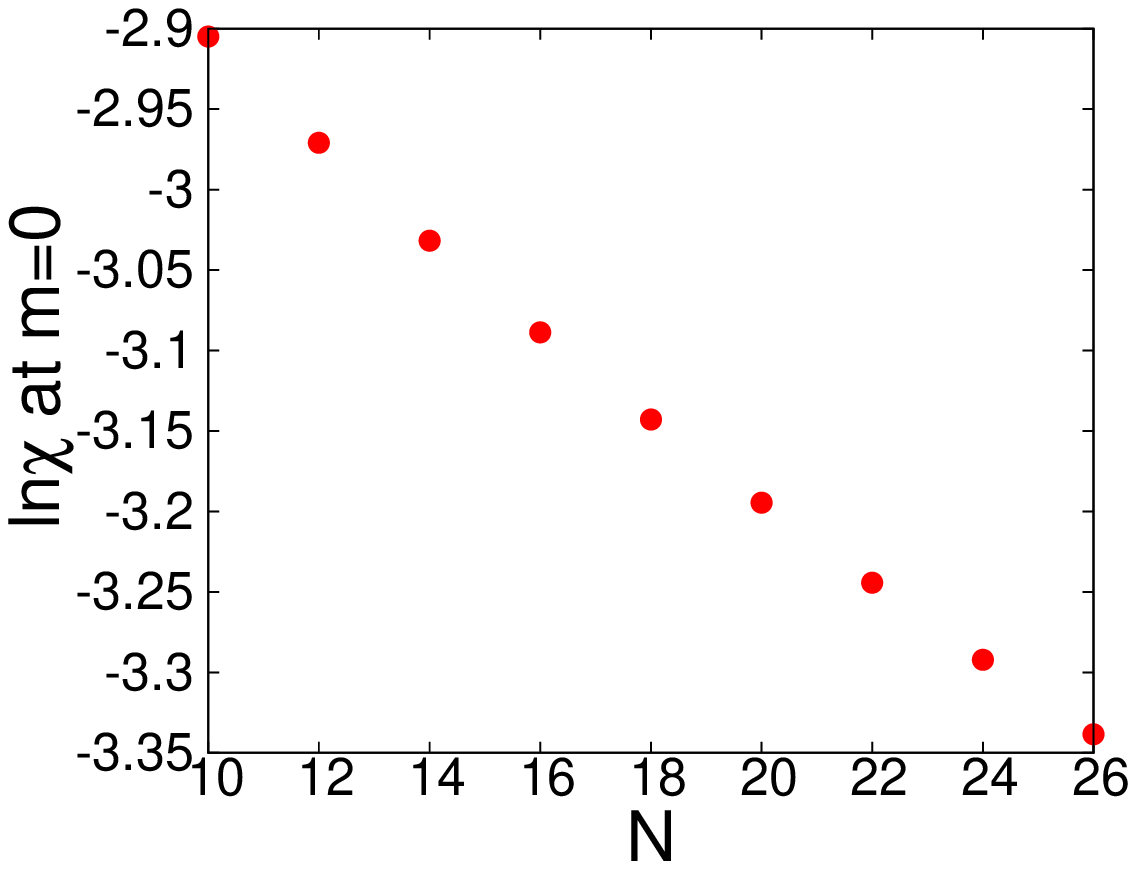}
  \subcaption{$\Delta=2$}
  \label{logchiN22}
\end{minipage}
\caption{
$N$ and $N^{-2}$ dependence of the magnetic susceptibility $\chi$ and its inverse 
$\chi^{-1}$ at zero magnetization.
Closed circles denote values of the magnetic susceptibility
for several system sizes $N$: 10, 12, 14, 16, 18, 20, 22, 24, and 26.
\subref{chiinN21} demonstrates that $\chi^{-1}$ 
becomes finite in the thermodynamic limit.
\subref{logchiN22} demonstrates that $\log{\chi}$ approaches minus infinity
in the thermodynamic limit, because $\log{\chi}$ is consistent with the
Ornstein--Zernike relation that explains $\ln{\chi} \propto -N/\xi + \ln{N}$,
where $\xi$ is the correlation length.
Therefore, \subref{chiN22} demonstrates that $\chi$ 
approaches zero in the thermodynamic limit and shows an anomaly that indicates
double degeneracy of ground states.
}
\label{chiinN2}
\end{figure*}

\subsection{scaling dimension and phase transition}
We explain that the scaling dimension $x_T$ is related to phase transition.
To demonstrate this, Fig.~\ref{4D0D} indicates the $\Delta$ dependence of the fourth derivative $A$ at zero magnetization. It appears that $A$ diverges for $\Delta>1/2$ as the size dependence is large.
This is as expected by C.N. Yang and C.P. Yang.\cite{5,31}
However, the anomaly of $A$ at $\Delta=1/2$ is different from that at $\Delta=1$ from the perspective of the scaling dimension $x_T$.
Generally, the $-1<\Delta\le1$ region corresponds to the Tommonaga--Luttinger (TL) phase, which is controlled by the scaling dimension $x_T$ subject to the parameter $\Delta$.\cite{5,22}
For $\Delta=1$, the anomaly of $A$ represents the phase transition corresponding to the KT transition when the scaling dimension $x_T$ changes from irrelevant to relevant for U(1) symmetry.
In contrast, for $\Delta=1/2$, the anomaly of $A$ does not represent the phase transition as the scaling dimension $x_T$ is irrelevant in the $-1<\Delta \le1$ region.\cite{45}
Similarly, the anomaly of $A$ for $\Delta=2$ does not represent the phase transition as the scaling dimension $x_T$ is relevant in $\Delta >1$.
Therefore, the phase transition is controlled by the scaling dimension $x_T$.
These facts show that the anomalies of $\chi$ and $A$ represent the phase transition through the scaling dimension $x_T$.
In addition to this, we name $-1<\Delta<1/2$ TL phase (I) and $1/2<\Delta\le1$ TL phase (I\hspace{-.1em}I) based on the behavior of $A$ for $\Delta>1/2$.

\subsection{Correction term and boundary conditions}
The correction term of Eq.~(\ref{7}) changes in relation to boundary conditions and dimension.
First, we discuss $C(N, m)$ in one-dimensional systems.
Without anomaly, $C(N, m)$ in a periodic boundary condition is written in the CFT as\cite{8,34,38}
\begin{align}
 C(N,m)= -\F{\pi v(m)}{6N^2},
\end{align}
where $v(m)$ is the velocity of the spin wave and a smooth function for $m$.
Thus, $\epsilon''(N,m)$ and $\epsilon^{(4)}(N,m)$ in Eq.~(\ref{nikai}) and Eq.~(\ref{yonkai})
converges to $1/N^2$ order, which agrees with our numerical results.
In contrast, the correction term for an open boundary is given by
\cite{8,34,38}
\begin{align}
 C(N,m)= \F{b(m)}{N}-\F{\pi v(m)}{24N^2},
\end{align}
where $b(m)$ is a non-universal boundary term.
In general, the convergence of this term is worse than that for a periodic
boundary condition.
We do not perform calculations for open boundary conditions herein,
and leave them for future work.

Next, we discuss the correction term in two-dimensional systems.
The correction term quickly converges, as shown by
Nakano and Sakai\cite{1}, and thus, has convergence of at least second order.
Unlike in the one-dimensional case, the convergence depends on the shape of the lattice.
Figure~4 in Ref.~\cite{1} is different from Fig.~\ref{chi}\subref{zika1}
from the perspective of an energy gap, although it resembles
Fig.~\ref{chi}\subref{zika1} from previous research.\cite{1}
This problem will be addressed in our future works.

\begin{figure*}[ht]
\begin{minipage}[t]{0.45\linewidth}
 \centering
  \includegraphics[keepaspectratio,scale=0.5]{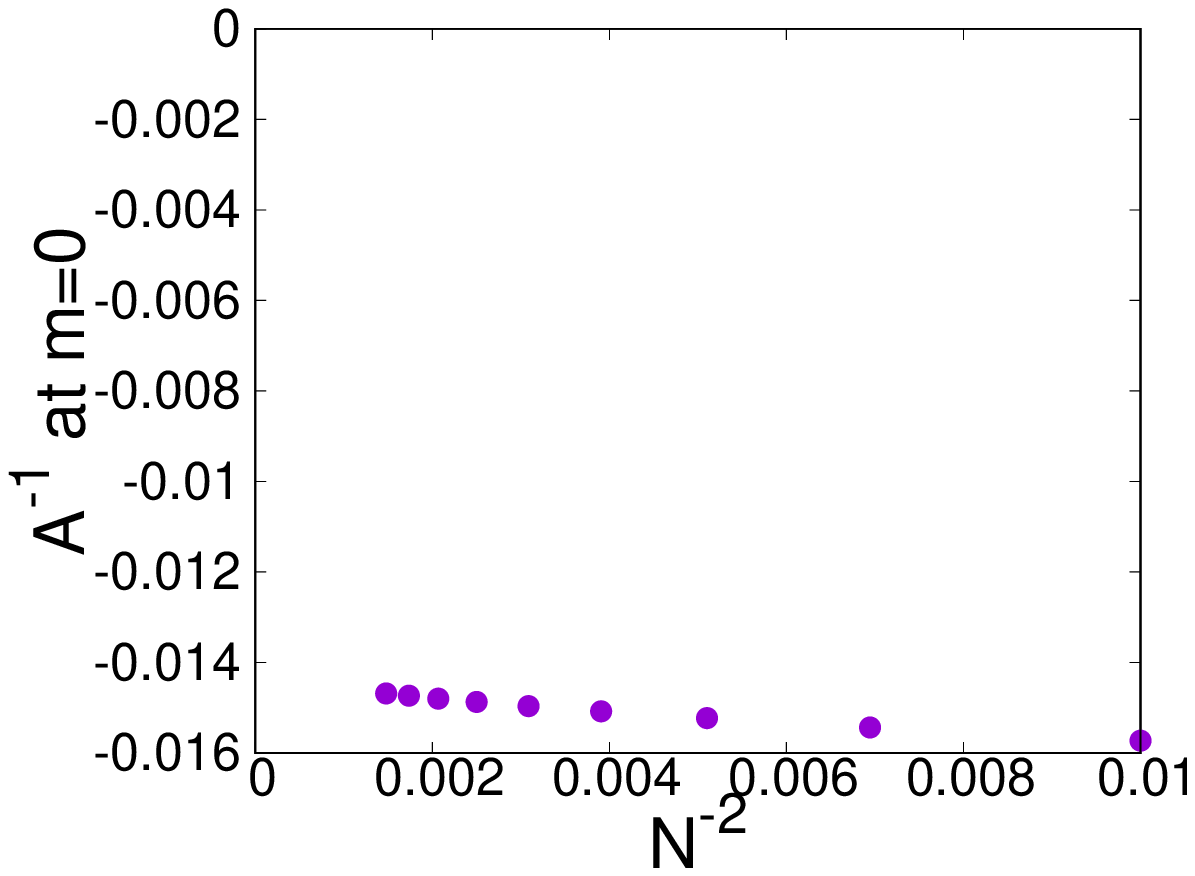}
 \subcaption{$\Delta=0.3$}\label{4DN203}
\end{minipage}
\begin{minipage}[t]{0.45\linewidth}
 \centering
  \includegraphics[keepaspectratio,scale=0.5]{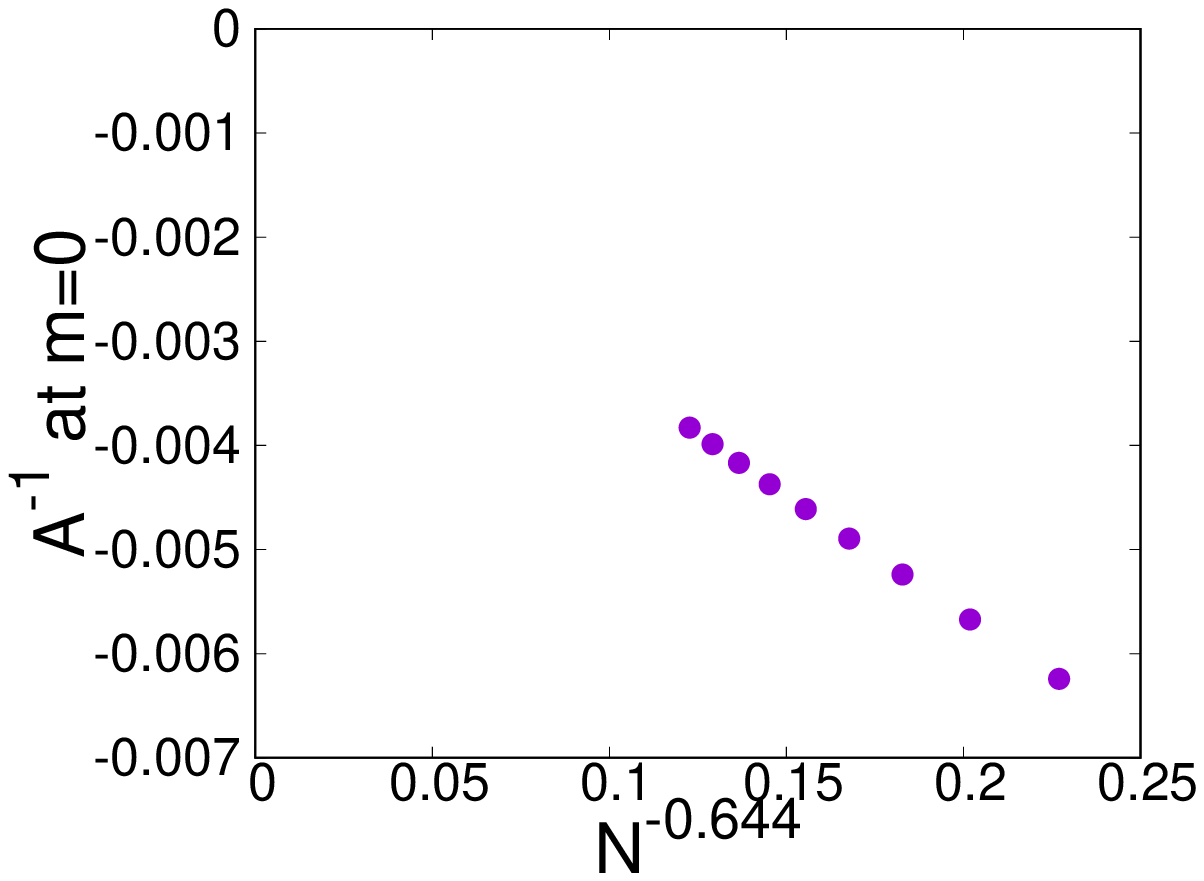}
  \subcaption{$\Delta=0.7$}\label{4DN06407}
\end{minipage}
\\
\begin{minipage}[t]{0.45\linewidth}
 \centering
  \includegraphics[keepaspectratio,scale=0.5]{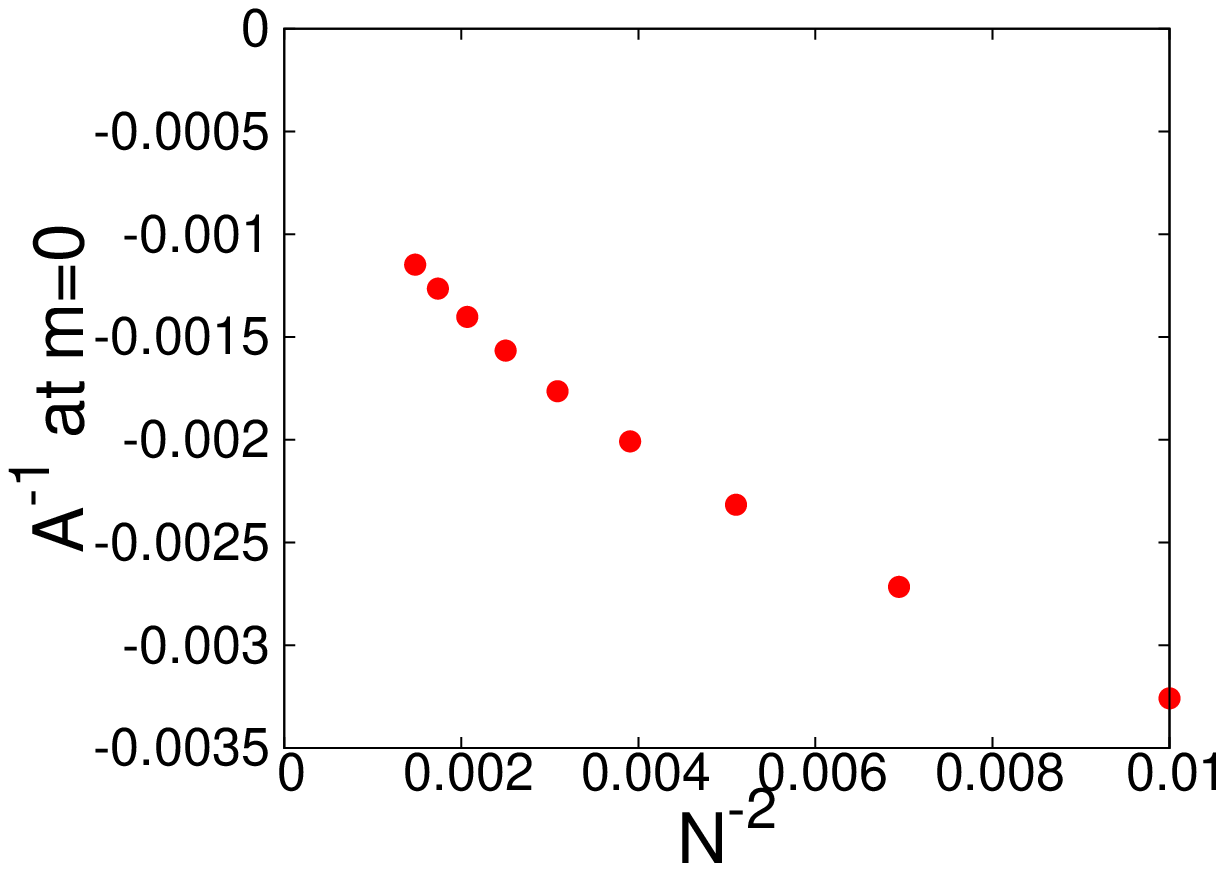}
 \subcaption{$\Delta=1$}\label{4DN21}
\end{minipage}
\begin{minipage}[t]{0.45\linewidth}
 \centering
  \includegraphics[keepaspectratio,scale=0.5]{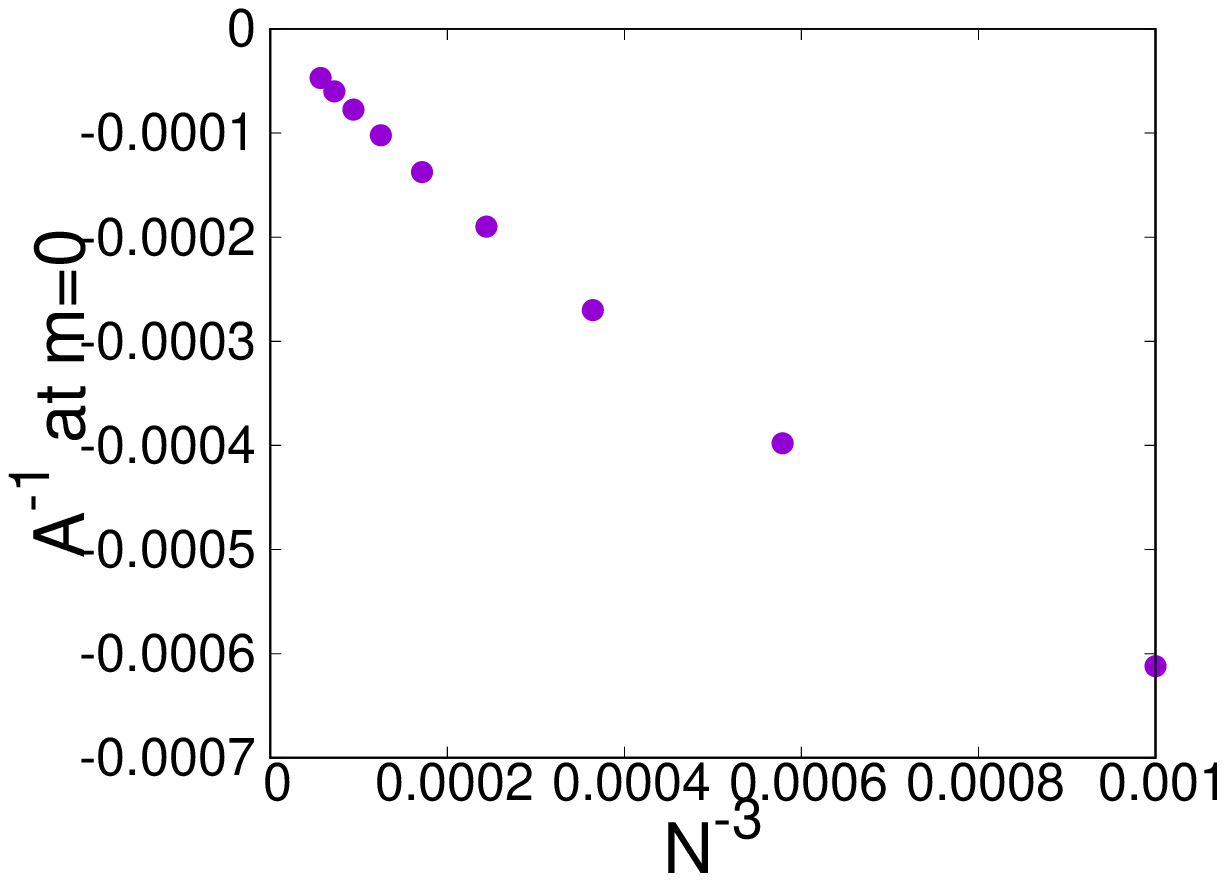}
  \subcaption{$\Delta=2$}\label{4DN22}
\end{minipage}
\caption{
$N$ dependence of the fourth derivative $A$ at zero magnetization.
Closed circles denote values of $A$ for several system sizes $N$:
10, 12, 14, 16, 18, 20, 22, 24, and 26.
\subref{4DN203} indicates that $A$ becomes constant in the thermodynamic limit.
\subref{4DN06407} indicates that $A$ approaches minus infinity and is consistent with Eq.~(\ref{yonkaiex}) in the thermodynamic limit.
Both \subref{4DN21} and \subref{4DN22} demonstrate that $A$ is minus infinity in the thermodynamic limit.
These numerical data are consistent with Eq.~(\ref{eq:yonkaixt2}) and Eq.~(\ref{eq:yonkaigap2}).
Thus, these graphs show an anomaly that indicates Kosterlitz--Thouless (KT)
transition for $\Delta=1$ and double degeneracy of ground states for $\Delta=2$.
}
\label{4DN2}
\end{figure*}

\begin{figure*}[ht]
\begin{minipage}[t]{0.45\linewidth}
 \centering
  \includegraphics[keepaspectratio,scale=0.5]{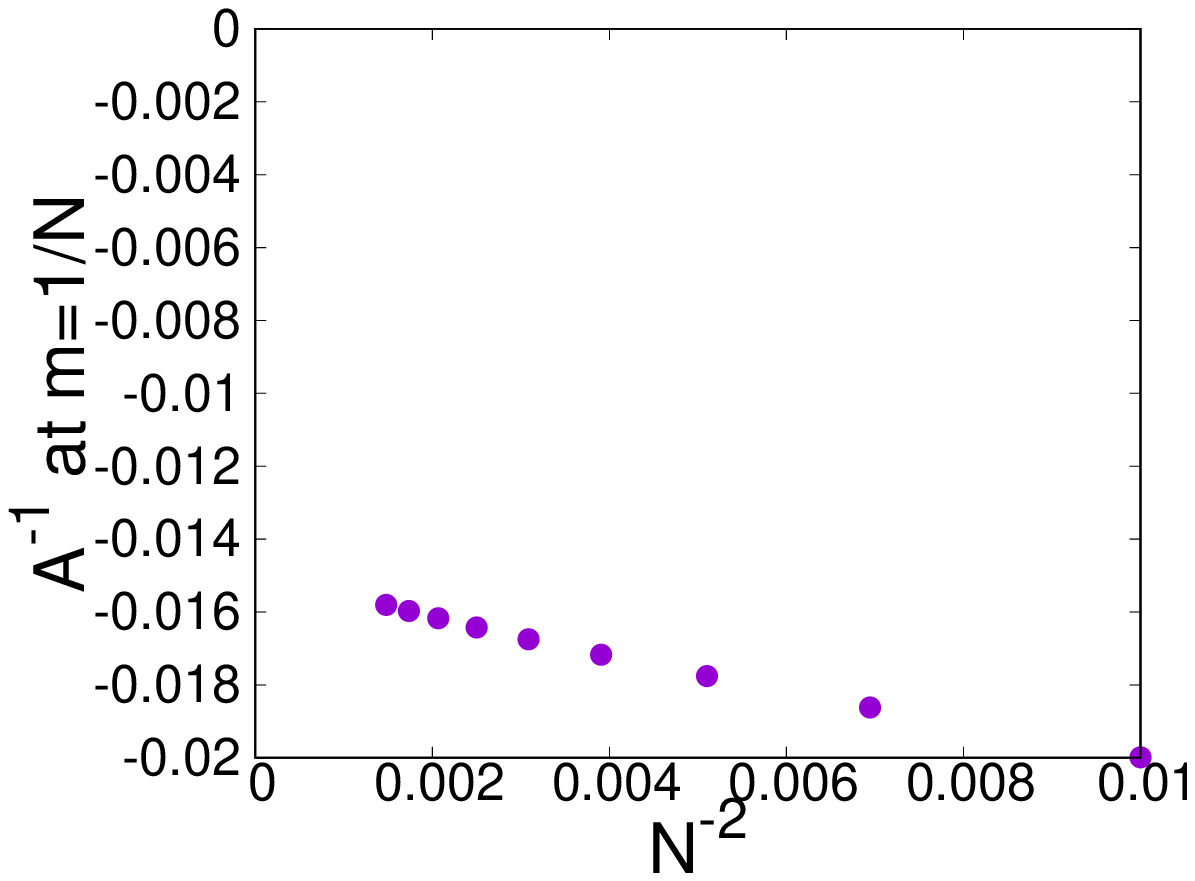}
 \subcaption{$\Delta=0.3$}\label{4DN203ap001}
\end{minipage}
\begin{minipage}[t]{0.45\linewidth}
 \centering
  \includegraphics[keepaspectratio,scale=0.5]{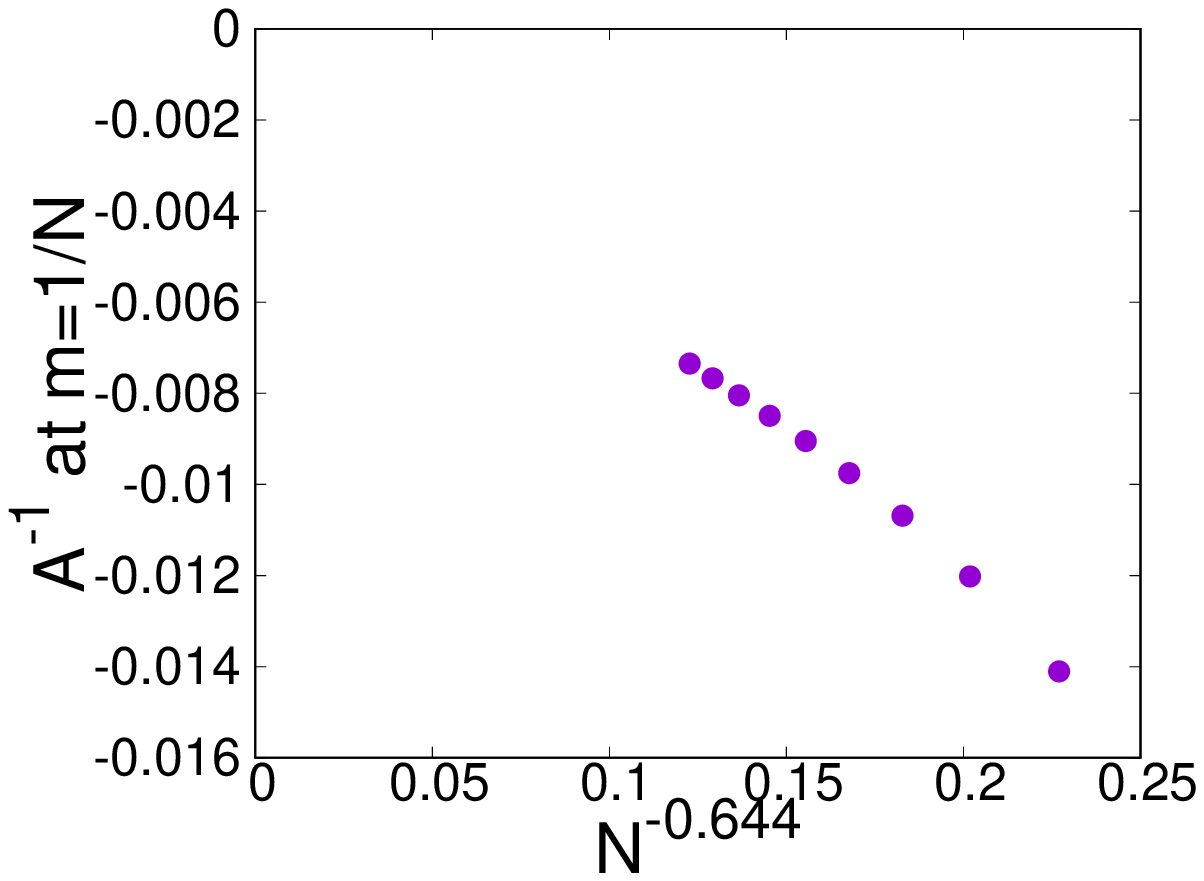}
 \subcaption{$\Delta=0.7$}\label{4DN06407ap001}
\end{minipage}\\
\begin{minipage}[t]{0.45\linewidth}
 \centering
  \includegraphics[keepaspectratio,scale=0.5]{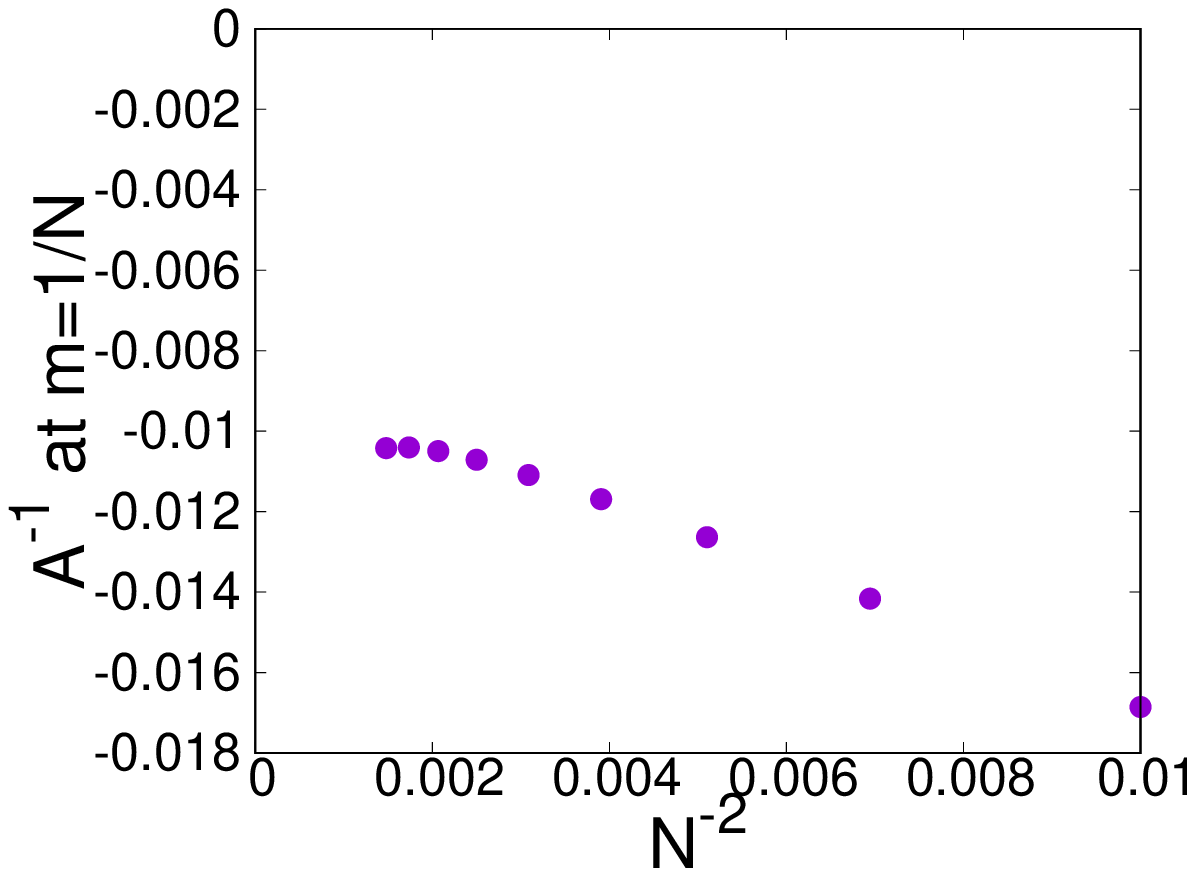}
 \subcaption{$\Delta=1$}\label{4DN21ap001}
\end{minipage}
\begin{minipage}[t]{0.45\linewidth}
 \centering
  \includegraphics[keepaspectratio,scale=0.5]{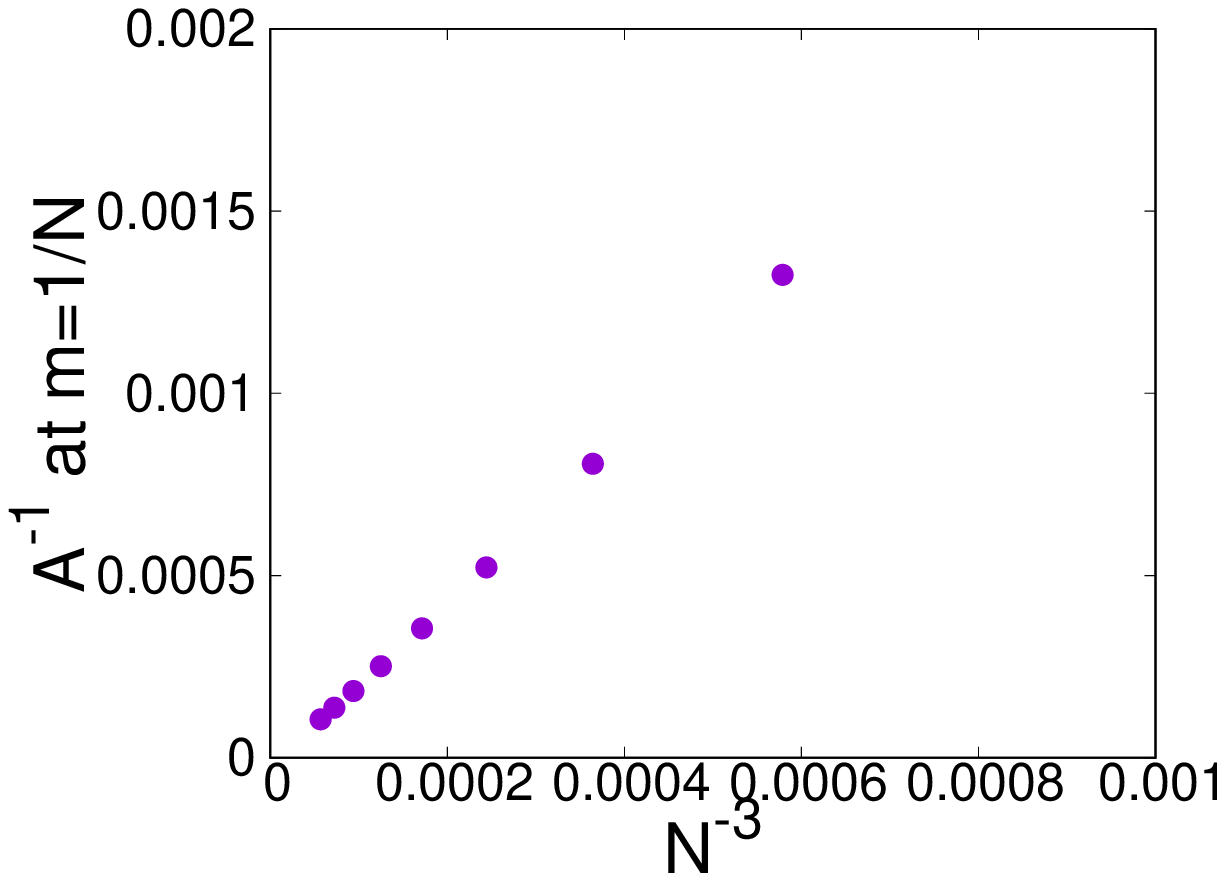}
 \subcaption{$\Delta=2$}\label{4DN22ap001}
\end{minipage}
 \caption{
$N$ dependence of the fourth derivative $A^{-1}$ at $m=1/N$.
Closed circles denote values of $A$ for several system sizes $N$:
10, 12, 14, 16, 18, 20, 22, 24, and 26.
\subref{4DN203ap001} indicates that $A$ becomes finite in the thermodynamic limit.
\subref{4DN06407ap001} indicates that $A$ approaches minus infinity and is consistent with Eq.~(\ref{yonkaicft2}) in the thermodynamic limit.
\subref{4DN21ap001} appears to indicate that $A$ becomes finite.
However, it disagrees with Eq.~(\ref{eq:yonkaixt2mnon0}), in which $A$ approaches infinity as the magnetization approaches zero. 
The disagreement results from the intermediate region in Eq.~(\ref{eq:yonkaixt2mnon0}), in which $A$ exhibits a negative and flat region.
Thus, $A$ approaches infinity as the system size becomes larger.
\subref{4DN22ap001} demonstrates that $A$ is infinity and is consistent
with Eq.~(\ref{eq:yonkaigap2}) in the thermodynamic limit.
Thus, \subref{4DN22ap001} shows an anomaly that indicates double degeneracy of ground states.
}
 \label{4DN2ap}
\end{figure*}

\section{Anomalies of $\chi^{-1}$ and $A$}
In this section, we investigate the anomalies of $\chi^{-1}$ and $A$
at zero and $m=1/N$ for $0\le \Delta \le2$
from the perspective of size dependence.
In addition, we reveal the anomalies of $\chi^{-1}$ and $A$.
The origin of an anomaly is usually a phase transition or Neel state that indicates double degeneracy of ground states with an energy gap for the $S=1/2$ XXZ chain.

First, the behaviors of $\chi$ and $\chi^{-1}$ are shown at zero magnetization in Fig.~\ref{chiinN2}.
Figure~\ref{chiinN2}\subref{chiinN21} shows that $\chi^{-1}$ becomes finite for $\Delta=1$ in the thermodynamic limit.
This indicates that the system does not have a finite spin gap or an anomaly.
In contrast, Fig.~\ref{chiinN2}\subref{chiN22} shows that $\chi^{-1}$
becomes infinite, i.e., $\chi$ reaches zero for $\Delta=2$ in the thermodynamic limit.
However, it is not conclusive that $\chi$ approaches zero in the thermodynamic limit.
To solve this problem, we present Fig.~\ref{chiinN2}\subref{logchiN22}, which is described as a semi-log graph of Fig.~\ref{chiinN2}\subref{chiN22}.
Figure~\ref{chiinN2}\subref{logchiN22} shows that
the behavior of $\log{\chi}$ is consistent with the
Ornstein--Zernike relation, which explains that $\ln{\chi} \propto -N/\xi + \ln{N}$,
where $\xi$ is the correlation length.
Thus, $\chi$ for $\Delta=2$ approaches zero, i.e., $\chi^{-1}$ approaches infinity in the thermodynamic limit.
This indicates that the system has a finite spin gap and an anomaly for $\Delta=2$.
The origin of the anomaly is the Neel state.
These facts are consistent with the results obtained by C.N. Yang and C.P. Yang;\cite{5,31} thus, we observed an anomaly of the magnetic susceptibility in a one-dimensional system. Moreover, the observation of the anomaly is useful for distinguishing gapped from gapless systems.

Next, the behavior of $A$ at zero magnetization is shown in Fig.~\ref{4DN2}.
Figure~\ref{4DN2}\subref{4DN203} shows that $A$ becomes finite for $\Delta=0.3$ in the thermodynamic limit, whereas Fig.~\ref{4DN2}\subref{4DN06407} shows that it is negative infinity for $\Delta=0.7$ in the thermodynamic limit.
For $\Delta=0.7$, from Eq.~(\ref{yonkaiex}), $A$ is proportional to $N^{6-2x_T}=N^{0.6439}$, as the scaling dimension $x_T=2.678003$ in Eq.~(\ref{scd}). Thus, we plot the horizontal axis in Fig.~\ref{4DN2}\subref{4DN06407} as $N^{-0.644}$.
The behavior of $A$ for $\Delta=0.7$ is consistent with Eq.~(\ref{yonkaiex}), in terms of both the power index and sign of the divergence.
Both Fig.~\ref{4DN2}\subref{4DN21} and Fig.~\ref{4DN2}\subref{4DN22} show that
$A$ is negative infinity in the thermodynamic limit.
For $\Delta=1$, from Eq.~(\ref{eq:yonkaixt2}), $A$ is proportional to $N^{6-2x_T}=N^{2}$, as the scaling dimension $x_T=2.0$. Thus, we plot the horizontal axis in Fig.~\ref{4DN2}\subref{4DN21} as $N^{-2}$.
The behavior of $A$ for $\Delta=1$ is consistent with Eq.~(\ref{eq:yonkaixt2}), in terms of both the power index and sign of the divergence.
For $\Delta=2$, from Eq.~(\ref{eq:yonkaigap2}), $A$ is proportional to $N^{3}$.
Therefore, we plot the horizontal axis in Fig.~\ref{4DN2}\subref{4DN22} as $N^{-3}$.
The behavior of $A$ for $\Delta=2$ is consistent with Eq.~(\ref{eq:yonkaigap2}), in terms of both the power index and sign of the divergence.
These demonstrate that $A$ shows an anomaly for $\Delta=0.7, 1.0, 2.0$.
However, the origin of the anomaly is different.
For $\Delta = 0.7$, the origin is TL phase (I\hspace{-.1em}I).
The origin of the anomaly for $\Delta=1$ is the phase transition, which means the transition from Tomonaga--Luttinger (TL) liquid phase to antiferromagnetic phase.\cite{5,22}
In contrast, in the $\Delta>1$ region, a Neel state appears for the $S$$=$$1/2$ XXZ chain. Thus, the origin of the anomaly for $\Delta=2$ is the Neel state.
Moreover, $A$ shows the transition for $\Delta=1$, although $\chi^{-1}$ does not show it 
from Fig.~\ref{chiinN2}\subref{chiinN21}.
The difference is used to confirm whether a phase transition happens or not. 
Therefore, observing $A$ is helpful for determining the consistency of phase transition.

Finally, we show $A$ at $m=1/N$ in Fig.~\ref{4DN2ap}, as the behavior of $A$ at $m=1/N$ differs between Fig.~\ref{A}\subref{A1} and Fig.~\ref{A}\subref{A2}.
Figure~\ref{4DN2ap}\subref{4DN203ap001} shows that $A$ becomes finite for $\Delta=0.3$ in the thermodynamic limit, whereas Fig.~\ref{4DN2ap}\subref{4DN06407} shows that it is minus infinity for $\Delta=0.7$ in the thermodynamic limit.
The behavior of $A$ for $\Delta=0.7$ is consistent with Eq.~(\ref{yonkaicft2}).
The origin of the anomaly is TL phase (I\hspace{-.1em}I).
It appears that $A$ in Fig.~\ref{4DN2ap}\subref{4DN21ap001} becomes finite for $\Delta=1$.
However, the behavior of $A$ is not consistent with Eq.~(\ref{eq:yonkaixt2mnon0}), in which $A$ reaches infinity when $m$ approaches zero.
The disagreement results from the intermediate region in Eq.~(\ref{eq:yonkaixt2mnon0}), in which $A$ exhibits flat and negative behavior, before reaching a sufficiently small region $|m| \ll 1$.
Thus, $A$ becomes infinite as the system size becomes larger in our calculation.
The origin of the anomaly is phase transition.
In contrast, Fig.~\ref{4DN2ap}\subref{4DN22ap001} shows that $A$ reaches infinity
for $\Delta=2$ in the thermodynamic limit.
The behavior of $A$ is consistent with Eq.~(\ref{eq:yonkaigap2}).
This indicates that $A$ shows an anomaly for $\Delta=2$.
The origin of the anomaly is a Neel state.
Therefore, the behavior of $A$ at $m=1/N$ in Fig.~\ref{A}\subref{A1} and Fig.~\ref{A}\subref{A2} is explained by Eq.~(\ref{eq:yonkaigap2}) and Eq.~(\ref{eq:yonkaixt2mnon0}).
Observation of the change in behavior at $m=1/N$ can be proposed as a new technique to
distinguish gapped from gapless systems.
Hence, observing $A$ at $m=1/N$ allows us to distinguish gapped from gapless systems.
However, future works must focus on the exact solutions of $A$ at $m=1/N$, as few investigations have focused on this behavior.

These findings indicate that observation of $A$ is more efficient than that of $\chi$.
Thus, we expect this technique to be used for analysis of spin liquids
with spin gap issues in triangular and Kagome lattices.\cite{1,28,29,30}

\section{Conclusion}
We investigated anomalies of $\chi$ and $A$ for the $S=1/2$ XXZ
antiferromagnetic chain by numerical diagonalization.
At zero magnetization, $\chi^{-1}$ shows an anomaly for $\Delta>1$. 
At zero magnetization, $A$ clearly indicates an anomaly for $\Delta>1/2$.
In addition, an anomaly of $A$ at $m=1/N$ is shown for $\Delta>1$.
In contrast, in the $\Delta<0$ region, future works are required regarding the anomalies of $\chi$ and $A$
in numerical calculations.
The results indicate that $\chi$ and $A$ have anomalies,
and that observing the anomaly of $A$ is easier than that of $\chi$
for relatively small system sizes.
In other words, the observation of phase transition is easier by $A$ than by $\chi$.
We reveal that the TL phase can be divided into $-1<\Delta<1/2$ as TL phase (I) and $1/2<\Delta\le1$ as TL phase (I\hspace{-.1em}I), from the perspective of the anomaly of $A$ at $\Delta=1/2$.
Therefore, we conclude that observation of $A$ is a useful method of
analyzing critical phenomena, compared with that of $\chi$.

Our study is concerned with one-dimensional systems.
However, our method can be used regardless of dimensions.
This method will help investigate quantum spin systems
in two or three dimensions.
In addition, this method can be applied to other systems such as spin liquids.
The behavior of spin liquids has been studied for magnetic susceptibility\cite{1,30}
and our method using $A$, compared with that using $\chi$, will be useful
for researching the behavior of a spin liquid that has spin gap issues.
The study of using $A$ for other models and higher dimension is left for future works.

In particular, $A$ relates to the nonlinear magnetic susceptibility\cite{25} of quantum spin systems and is thus of direct relevance to experiments.
The nonlinear magnetic susceptibility can be easily calculated
with high accuracy using $A$.
The method using $A$ can be a new technique in the study of quantum spin systems
and strongly correlated electron systems.
Furthermore, the method will enable one to discover a magnetization plateau observed in experiments that shows constant magnetization when a magnetic field changes.
The plateau indicates the anomaly of $\chi$, and that of $A$ should also appear there
from Eq.(\ref{eq:chigap2}), (\ref{eq:yonkaigap2}).
The observations of $\chi$ and $A$ will be useful for evaluating
a magnetization plateau.
Numerical diagonalization calculations of $A$ will provide us with
a new development in theory and experiments for quantum spin systems.
\section{Acknowledgment}
We would like to thank Professors T. Sakai, M. Takahashi, T. Matsui, and J. Fukuda
for their helpful discussions.
We would like to thank Editage (www.editage.com) for English language editing.
Our calculations on numerical diagonalization were performed using TITPACK Ver.2,
which Professor H. Nishimori coded, and $H\phi$, which Professor M. Kawamura {\it et al.} coded.

\bibliography{ref}
\nocite{*}

\end{document}